\documentclass[12pt]{article}

\usepackage{scicite}
\usepackage{amsmath}
\usepackage{graphicx}
\usepackage{float}
\usepackage[titletoc]{appendix}
\usepackage{epstopdf}
\usepackage{color}
\usepackage{times}
\usepackage[dvipsnames]{xcolor}
\usepackage{soul}
\usepackage{url}

\newcommand{\br}[1]{\left( #1 \right)}

\newcommand{\Ro}{\textrm{Ro}}

\newcommand{\bu}{\boldsymbol{u}}

\newcommand{\bx}{\boldsymbol{x}}

\newcommand{\bF}{\boldsymbol{F}}
\newcommand{\bD}{\boldsymbol{D}}

\newcommand{\bV}{\boldsymbol{V}}

\newcommand{\bor}{\boldsymbol{r}}

\newcommand{\pa}[1]{\partial_{#1}}

\newcommand{\ovl}[1]{\overline{#1}}

\newenvironment{sciabstract}{%
\begin{quote} \bf}
{\end{quote}}

\title{Direct observational evidence of an oceanic dual kinetic energy cascade and its seasonality}

\author
{Dhruv Balwada$^{1\ddagger}$, Jin-Han Xie$^{2\ddagger \ast}$, Raffaele Marino$^3$, and Fabio Feraco$^3$\\
\\
\normalsize{$^{1}$Lamont-Doherty Earth Observatory, Columbia University, New York, NY, USA.}\\
\normalsize{$^{2}$Department of Mechanics and Engineering Science at College of Engineering and LTCS,}\\
\normalsize{Peking University, Beijing, PR China.}\\
\normalsize{$^{3}$ Univ Lyon, CNRS, \'Ecole Centrale de Lyon, INSA Lyon, Univ Claude Bernard Lyon 1,}\\
\normalsize{Laboratoire de M\'ecanique des Fluides et d'Acoustique,  UMR5509, F-69134 \'Ecully, France.}\\
\normalsize{$^\ddagger$ These authors contributed equally to this work.}\\
\normalsize{$^\ast$Corresponding author. Email: jinhanxie@pku.edu.cn.}
}

\date{}

\begin{document} 
\baselineskip24pt
\maketitle 


% Guide to writing structuring: https://journals.plos.org/ploscompbiol/article?id=10.1371/journal.pcbi.1005619

% Guide to write abstract/summary paragraph: https://www.nature.com/documents/nature-summary-paragraph.pdf

% List of important papers related to ocean seasonality and Gulf of Mexico: https://docs.google.com/document/d/1QU66A6qROuow4Z87QlygZF82xCAfzXR0jus0gO3QCJY/edit?usp=sharing

%%%%%%%%% ABSTRACT %%%%%%%%%%%
\begin{sciabstract}

The Ocean’s turbulent energy cycle has a paradox; large-scale eddies under the control of Earth’s rotation primarily transfer kinetic energy (KE) to larger scales via an inverse cascade, while a transfer to smaller scales is needed to accomplish dissipation. It has been argued, using numerical simulations, that fronts, waves and other turbulent structures can produce a forward cascade of KE toward dissipation scales.  However, this forward cascade and its coexistence with known inverse cascade were not confirmed in observations. Here we present the first evidence of a dual KE cascade in the Ocean by analyzing velocity measurements from surface drifters released in the Gulf of Mexico. Our results show that KE is injected at two dominant scales and transferred to both large and small scales, with the downscale flux dominating at scales smaller than ~1-10km. The cascade rates are modulated seasonally, with stronger KE injection and forward transfer during winter.

\end{sciabstract}

% In setting up this template for *Science* papers, we've used both
% the \section* command and the \paragraph* command for topical
% divisions.  Which you use will of course depend on the type of paper
% you're writing.  Review Articles tend to have displayed headings, for
% which \section* is more appropriate; Research Articles, when they have
% formal topical divisions at all, tend to signal them with bold text
% that runs into the paragraph, for which \paragraph* is the right
% choice.  Either way, use the asterisk (*) modifier, as shown, to

%%%%%%%%%%%%%%%%%%%%%%%%%%%%%%%%%
%%%%%%%% Introduction %%%%%%%%%%%
%%%%%%%%%%%%%%%%%%%%%%%%%%%%%%%%%
%\section*{Introduction}

%% Para 1
% - Why interscale transfers matter in the ocean. 
% - some basics of what is known about these 
% - Our knowledge of downscale transfers is still in infancy 
The oceanic circulation is primarily forced at scales of $O(1000)$ km by the winds, tides, and solar heating, and dissipated by friction at scales of $O(1)$~mm. Ocean turbulence helps to redistribute the energy across scales, populating the range between forcing and dissipation scales, and also promoting the exchanges between potential and kinetic energy reservoirs \cite{ferrari2009ocean}. 
%\textit{In rotating stratified flows, e.g. as described by interior quasi-geostrophic (QG) dynamics, the general tendency of the flow is to transfer kinetic energy from small to large scales, through the inverse cascade \cite{Salmon1982, vallis2017atmospheric}, and generate a rather steep drop-off in kinetic energy at the smaller scales, following a $k^{-3}$ spectra. 
%	These dynamics suggest that the bulk of the kinetic energy would reside in the mesoscale eddies $O(100)$ km, as is the case in the ocean \cite{tulloch2011scales}, and that this energy will be dissipated at the large scales, which is not easily physically justifiable. It is expected that dissipation will be active at the smallest scales, and thus mechanisms not described in QG theory have to be active for transferring kinetic energy from large to small scales, through the forward cascade.}
The bulk of the oceanic kinetic energy resides in flows with horizontal scales of hundreds of kilometers \cite{tulloch2011scales}, the so-called mesoscales. These flows are characterized by rapid rotation and strong stratification, so their dynamics are well described by the quasi-geostrophic (QG) theory. These dynamics dictate a general tendency of the oceanic flow to transfer kinetic energy from small to large scales through the inverse cascade \cite{Salmon1982, vallis2017atmospheric}, and suggest a rather steep drop-off in kinetic energy at the smaller scales, following a $k^{-3}$ spectrum. Consequently, according to QG dynamics, the mesoscale kinetic energy is expected to be dissipated primarily by boundary friction. However, estimates suggest that dissipation through boundary friction accounts for only one-tenth of the total energy injection, raising a puzzle about the dissipation mechanisms \cite{Wunsch2004}. To resolve this puzzle, there must be mechanisms not described in the QG theory for transferring kinetic energy from large to small scales, through a forward cascade.

Observations and high-resolution simulations have shown that the range of scales referred to as the submesoscales, scales between 3D turbulence (O(50-100)m) and the mesoscale (O (100m)), are quite energetic in the surface ocean \cite{Callies2014, rocha2016mesoscale}, often following a $k^{-2}$ spectral slope for horizontal kinetic energy and buoyancy variance, and show a pronounced seasonal modulation in energy levels \cite{sasaki2014impact, callies2015seasonality, buckingham2016seasonality,Zhang2021}. 
These scales are thought to be energized through mesoscale driven straining of buoyancy fronts \cite{Lapeyre2006,Roullet2012} or mixed-layer instabilities \cite{Boccalett2007,Callies2016}, which acts to release the available potential energy stored in mixed layers. The latter mechanism is now routinely implicated for the observed seasonality at these scales. 
The shallow spectral slope implies that the Rossby number, a ratio of inertial force to the Coriollis force, can become O(1) at these scales, which suggests that the submesoscale flows can escape the constraints of geostrophic balance and potentially transfer kinetic energy to smaller scales \cite{mcwilliams2016submesoscale, klein2019ocean}. These flows are also expected to play an important role in the mixed layer restratification \cite{su2018ocean}, and transporting tracers between the mixed layer and the interior \cite{balwada2018submesoscale, uchida2019contribution}. 

The suggestion that submesoscale flows at the surface can result in a forward cascade of kinetic energy has been confirmed in high-resolution ocean model runs, e.g. \cite{capet2008mesoscale, yang2020spatial}.  These simulations show that the surface kinetic energy flux can undergo a dual cascade, flowing upscale at large scales and downscale at small scales, with the forward cascade being present at scales roughly smaller than O(10)km. The kinetic energy reservoir is supplied by conversion from available potential energy to kinetic energy, and the ageostrophic flow is crucial for the forward cascade to emerge. In this phenomenology the surface kinetic energy does not develop an inertial range, as it is not a conservative quantity because of exchanges with the underlying interior \cite{mcwilliams2016submesoscale}. Additionally, direct numerical simulations of rotating stratified turbulent flows in the appropriate parameter regimes \cite{Pouquet2017, Marino2013} have indicated that a dual energy cascade can also be sustained in the ocean interior. While evidence and mechanistic understanding of the dual cascade in the ocean has been made possible by sophisticated high-resolution simulations, there are no observational studies that have been able to unambiguously confirm its presence yet. 

Estimating the inter-scale energy transfers in the ocean from observations is extremely challenging, due to the fact that conventional spectral flux estimation methods used to analyze numerical simulations \cite{Frisch1995} require the availability of synoptic measurements on a regularly sampled grid and over a fairly large region; technology to collect such measurements at submesoscales is not available at this time \cite{klein2019ocean}. 
Sea surface height (SSH) measurements from satellites come the closest to producing datasets that are amenable to spectral flux calculations, they provide gridded estimates of surface geostrophic velocity with a nominal spatial resolution of O(100)km and time resolution of a week. Spectral flux calculations from SSH-based velocities have provided clear evidence for the presence of an inverse energy cascade at scales larger than 100km, and suggested the presence of a forward energy and enstrophy cascade at scales smaller than 100km \cite{Scott2005, khatri2018surface}. However, the conclusions about scales smaller than 100km are debatable because these estimates rely on the assumption of geostrophy to estimate the velocity from SSH and due to strong sensitivity to the gridding and interpolation methods\cite{Aluie2018}. 
In coastal locations, high-frequency radars have been used to estimate the surface velocity fields with resolutions of a few kilometers and can be used for estimating the spectral flux, but are limited in coverage to within $\sim 10-100$km off the coast \cite{harlan2010integrated}. 
Velocity estimates from a mooring array were used in a recent study to suggest the presence of a forward cascade at the submesoscales in the spring \cite{garabato2022kinetic} using a frequency decomposition. However, this work relied on assuming a connection between the temporal and spatial scales due, and did not directly probe the spatial structure of how kinetic energy is transferred.

An approach to probing the properties of the inter-scale energy transfers using observations has been to use third-order structure functions \cite{Cho2001, Balwada2016}, which can be estimated from un-gridded or scattered measurements under the assumption of statistical homogeneity.
Surface drifters, released in large clusters, have allowed for a characterization of the energy cascades by estimating third-order velocity structure functions (SF3, see Methods) \cite{Balwada2016, poje2017evidence, berta2020submesoscale}. These studies, all from drifter releases in the Northern Gulf of Mexico, have suggested that a forward cascade exists at scales smaller than O(1-10)km based on a sign reversal in SF3 around this scale, and have quantified the forward energy flux based on exact formulae for the SF3 derived using inertial-range arguments. 
%However, most theoretical work surrounding the SF3 and estimating the energy flux from them have relied on inertial range assumptions \cite{Frisch1995}. 
%Since we are not sure about the existence of an inertial range for the surface KE, any quantification of the energy flux using the inertial-range theories is suspect.
%Also, inertial range theories assume that the spectral fluxes are unidirectional over a range of scales, and thus are quite limited for analysis of flows that likely portray dual cascades.
A major improvement of this methodology for the quantification of the energy fluxes is due to Xie and B\"uhler 2019 \cite{XieBuhler2019b}, who proposed a forcing-scale resolving SF3 formulation able to capture the simultaneous bidirectional energy transfer. 
This formulation applies to a range beyond the inertial range, and its implementation does not require the identification of inertial ranges in order to fit SF3 expressions with power-law functions.
Meanwhile, it also allows the estimation of energy injection scales, which is important for gaining a better understanding of the turbulent cascades.

%Also, the new theory naturally expresses different inertial ranges in one formula, which enables us to convincingly detect bidirectional energy transfer.

Here we apply these new theoretical insights to two surface drifter data sets, collected during the Grand Lagrangian Deployment (GLAD) and the Lagrangian Submesoscale Experiment (LASER) in the Gulf of Mexico in summer and winter, respectively. We focus specifically on velocity estimated from surface drifter trajectories in the northern Gulf of Mexico, in waters that are deeper than 500m and away from the continental shelf.
Our results confirm the presence of a seasonal modulation of submesoscale surface kinetic energy (KE), and more importantly, for the first time, characterize and quantify the inter-scale energy transfer simulataneously to the large and small scales, at the submesoscales in the ocean.
The results confirm the existence of a dual cascade of KE at the ocean surface, which is energized primarily at scales close to the mixed-layer and interior deformation radii, with the transition of energy flux direction happening around a scale where the local Rossby number is of O(1).
%\textit{There is a stronger injection and flux of KE during the winter, and the scale at which this stronger injection takes place moves to larger scales during the winter.} 
Energy injection and the KE flux are stronger in the winter than in the summer, and the scale of energy-injection likely corresponding to the mixed layer baroclinic instability also moves to larger scales in the winter relative to the summer.

\section*{Results}
\subsection*{GLAD and LASER experiments}
\begin{figure}
	\centering
	\includegraphics[width=1\linewidth]{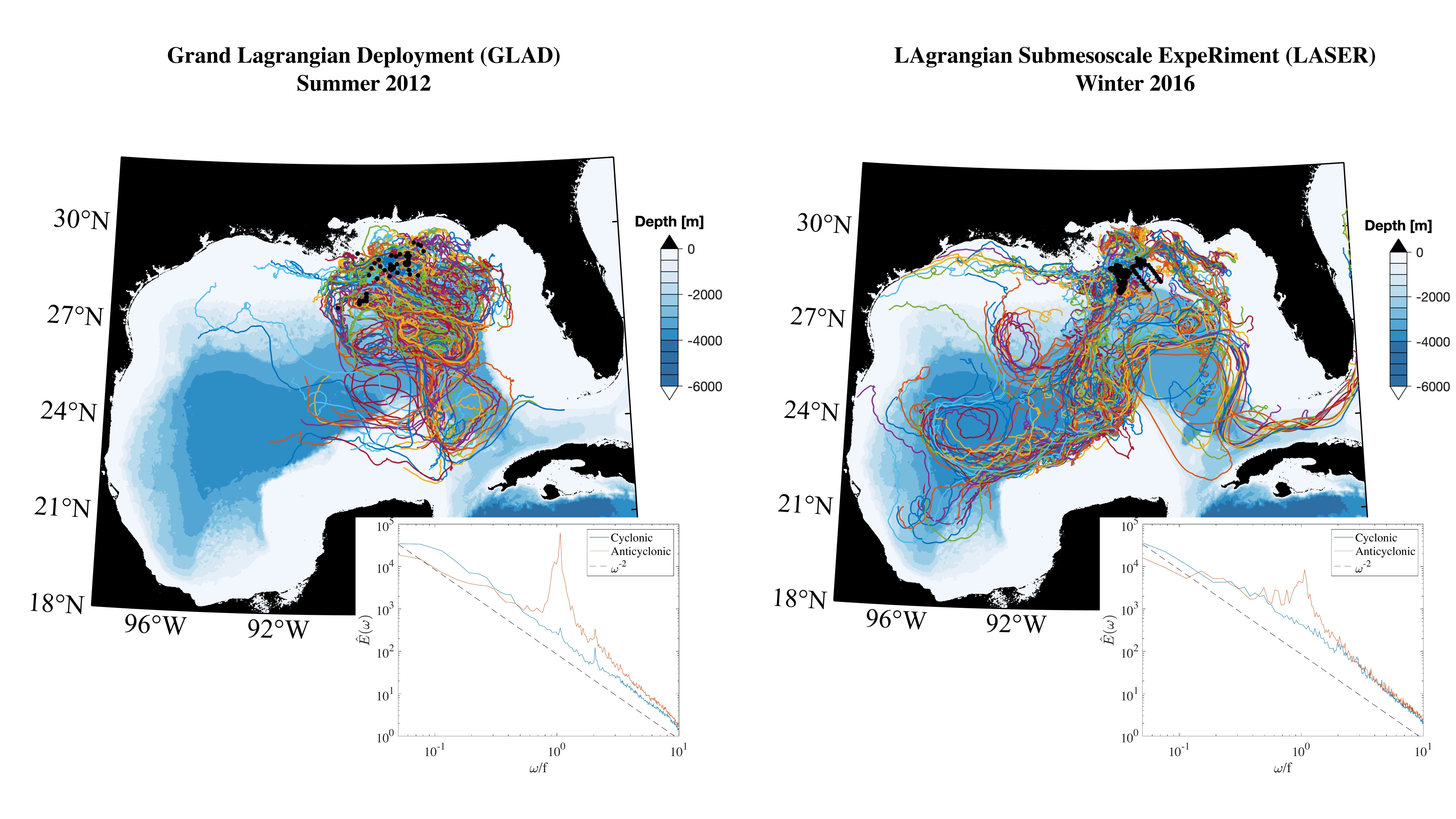}
	\caption{\textbf{Spatial distribution of the drifters.} Drifter tracks and the Lagrangian frequency spectra (inset plots) for the GLAD (a) and LASER (b) experiments. Black dots indicate the deployment locations, and the colored contours indicate the bathymetric depth. In our study we only considered the sections of the drifter tracks that were in water deeper than 500m, and in between east of 84-91$^o$W and north of 24$^o$N.}
	\label{fig:tracks}
\end{figure}

% data summary
We analyze data from surface drifters that were deployed in the Northern Gulf of Mexico, in a region usually to the north of the Loop Current. The surface drifters were deployed as part of the Grand Lagrangian Deployment (GLAD) in Summer/July-August 2012 \cite{poje2014submesoscale} and the Lagrangian Submesoscale Experiment in Winter/January-February 2016 \cite{dasaro2018ocean}, which are to-date the largest simultaneous drifter deployments. Over the course of the 3-4 months that the drifters were active they dispersed to span a large part of the Gulf of Mexico, and this long-term dispersion was largely influenced by the basin-scale and mesoscale circulation in the region (Figure \ref{fig:tracks}). The part of the data analyzed here comes mainly from the initial few weeks after the deployments when the drifters are relatively close to each other, and contained mostly in the Northern Gulf. We also excluded the drifter tracks when they went in waters shallower than 500m or ventured west of 91$^o$W, east of 84$^o$W, or south of 24$^o$N, as we want to focus on the dynamics away from the continental shelf and in the NorthEastern GOM.

% regional characterization and seasonal differences
The atmospheric forcing during summer deployment was characterized by relatively weak  winds ($\sim 5$ m/s), while the winter deployment experienced stronger  winds ($\sim 8$ m/s) and  severe storms. The summer months in the northern GOM are also characterized by very shallow mixed layers ($\sim 10$m) and lateral buoyancy gradients that are produced by the inflow of fresh water from the Mississippi river delta, while in the winter the mixed layer deepens ($\sim 80$m) and the lateral buoyancy gradients are primarily a result of temperature variations \cite{berta2020submesoscale, barkan2017submesoscale}. The drifters tracked the flow in the top 0.60 m of the water column \cite{novelli2017biodegradable}. The summer drifter trajectories and velocities show a marked presence of inertial oscillations (Figure \ref{fig:tracks} inset), and the amplitude of these oscillations are damped by about an order of magnitude in the winter; likely in response to the seasonal modulation of the mixed layer depth \cite{d1978mixed}. 

% structure functions 
%Here we will consider the second and third order velocity structure functions from the drifters to infer the multi-scale properties of the surface kinetic energy field, in particular its distribution and inter-scale transfers (see methods for details).
The scale-wise variation of the kinetic energy content in the surface ocean will be assessed by means of the second-order velocity structure functions, and then a novel approach based on the third-order structure functions will be implemented to get insights in to the transfer and injection of kinetic energy as a function of length scales (see method for details).

\subsection*{Seasonal modulation of surface kinetic energy}

\begin{figure}
	\centering
	\includegraphics[width=1\linewidth]{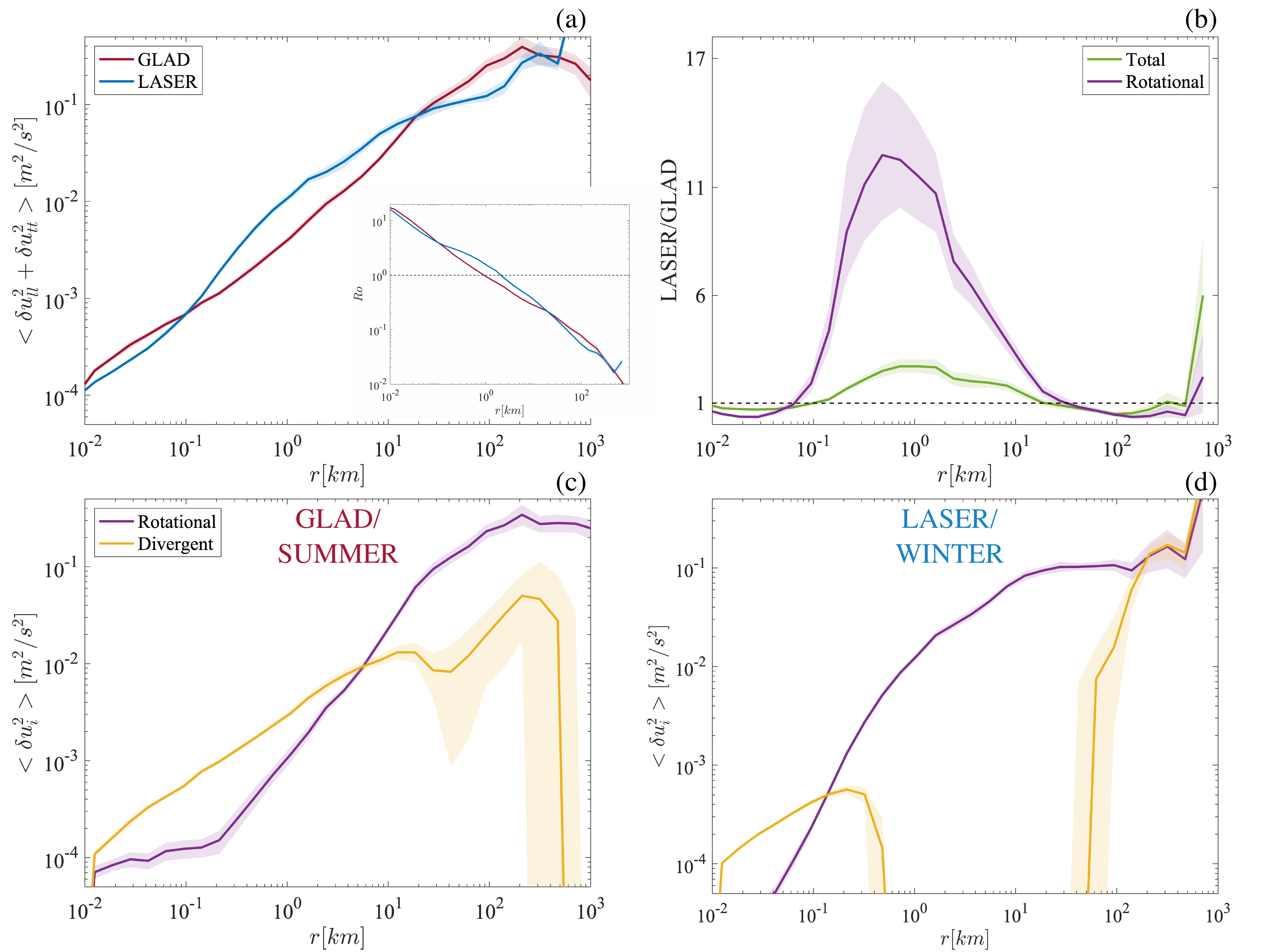}
	\caption{\textbf{Distribution of kinetic energy as a function of scale represented using second-order structure functions.} (a) The total, sum of the longitudinal and transverse, second-order structure function ($D_{tot}(r) = D_{LL}(r) + D_{TT}(r)$) as a function of separation scale for the GLAD and LASER experiments. The inset shows the Rossby number, defined as $Ro=\sqrt{D_{tot}}/fr$, as a function of scale, with the dashed horizontal line indicating $Ro=1$. (b) The ratio of total and rotational second order structure function from the LASER and GLAD experiments, with dashed horizontal line indicating a ratio of 1. (c, d) The decomposition of the total structure function into the rotational and divergent components, Helmholtz decomposition, for the GLAD and LASER experiments.  }
	\label{fig:S2}
\end{figure}

The second-order structure functions (SF2; $D_{LL} (r) = \left< \delta u_L^2 \right>,  D_{TT} (r) = \left< \delta u_T^2 \right>$), which reflect how kinetic energy is distributed as a function of scales, show a marked change in properties between summer and winter (Figure \ref{fig:S2})\footnote{Some aspects of the second order structure functions presented here have been  discussed for the individual experiments or for a shorter range of scales in previous studies, eg \cite{poje2014submesoscale, Balwada2016, poje2017evidence, berta2020submesoscale, wang2021anisotropic}, but this is the first comparative synopsis between seasons using the Helmholtz decomposition.}. 
The detailed definition of SF2 can be found in the Method section (cf. (\ref{SF2_def})).
The total SF2 ($D_{tot} = D_{LL} + D_{TT}$), reflective of the total horizontal kinetic energy, in winter is larger by a factor of approximately 2 at scales on the O(100m-10km), while the non-divergent or rotational part of SF2 (discussed more in the next paragraph) in winter is around 10 times larger at those scales (Figure \ref{fig:S2}b). The larger scales, O(20-100km), show a slight reduction of total SF2 in the winter; this is most likely a reflection of the synoptic modulation in the mesoscale field rather than a signal associated with the seasonal variability. The smallest scales, O($<$100m), also show a slight reduction of total SF2 in winter, which is likely related to the precise deployment conditions and spatial variability since these smallest range of scales are sampled for very short periods after deployment. We also defined a scale-dependent Rossby number using the total SF2, as $Ro = \sqrt{D_{tot}(r)} /fr$ with $f$ being the Coriolis frequency and $r$ being the separation scale. This Rossby number is O(1) at scales smaller than 1-5~km, and is slightly greater in the winter than the summer following the seasonal modulation of the total SF2. 

To gain further insight into the type of the flows that contribute to the total SF2, we decomposed it into the rotational ($D_{R}$) and divergent ($D_{D}$) contributions using a Helmholtz decomposition \cite{Lindborg2015}. 
The relative contribution from the divergent and rotational parts of the flow  changes significantly between seasons, and indicates a dramatic change in the flow dynamics between summer and winter (Figure \ref{fig:S2}c,d). In summer the divergent motions dominate up to scales of 5km, while in winter dominance of divergent motions is limited to scales smaller than about 100m. Non-physical peculiarities in the divergent component, such as small negative values (at scales larger than 500km in summer and scales between 500m-50km in winter) or comparable values at the largest scales (scales larger than 200km in winter) are likely a result of slight breakdown of assumptions of homogeneity and isotropy that are required for the Helmholtz decomposition \cite{Buhler2014}; we do not expect these issues to impact the qualitative aspects of these results \cite{wang2021anisotropic}. 
The power-law behavior of the rotational SF2 also shows a marked change between seasons. While both seasons show that the rotational SF2 has a shallow slope, suggestive of an inverse energy cascade, at large scales and steeper slope at smaller scales, the scale where this slope changes shifts from approximately 1km in winter to around 20km in summer. 

The second-order structure functions indicate that the submesoscale flow, roughly defined as scales smaller than 50km, is more energetic in winter than the summer, and this is primarily a result of strengthening of the rotational (non-divergent) flow. 
The change in power-law behavior of the rotational SF2 between seasons suggests that a higher amount of kinetic energy is injected into the non-divergent part of the flow near the mixed-layer deformation radius, O(1-10km), in the winter, and the cascade of this energy is strong enough to paint the distribution of kinetic energy in the submesoscales. 
It is also quite striking that the energized winter flow is dominantly rotational, which suggests a dominance of balanced flows likely energized by mixed-layer instability, while the summer flow has a large contribution from divergent modes suggesting an abundance of ageostrophic motions, potentially resulting from strain-driven frontogenesis and internal waves (also highlighted in the Lagrangian frequency spectra in Figure \ref{fig:tracks}). 

The results from the surface drifters suggest that the seasonality of the surface flows in this part of the Gulf of Mexico are qualitatively similar to other parts of the ocean that have a large seasonal modulation of mixed-layer depth \cite{Callies2016, rocha2016mesoscale}. 

\subsection*{Seasonality of inter-scale kinetic energy transfers}

\begin{figure}
	\centering
	\includegraphics[width=0.7\linewidth]{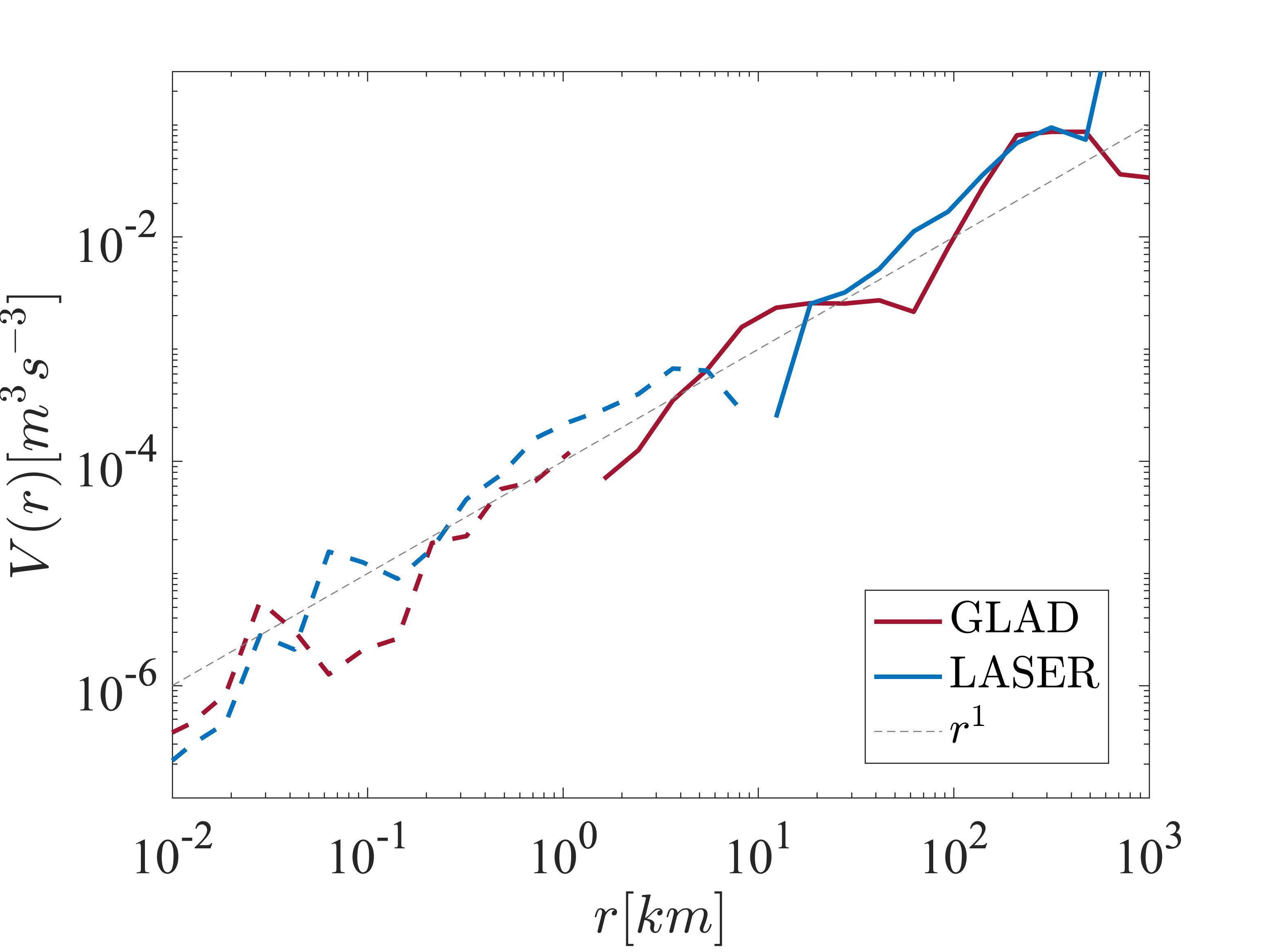}
	\caption{\textbf{Third-order structure functions ($V(r) = \left< \delta u_L^3\right> + \left< \delta u_L \delta u_T^2 \right> $) from GLAD and LASER.} Absolute value of $V(r)$ is plotted, and the range of scales where $V(r)$ is negative are indicated as dashed lines. A linear power law is also indicated as a dashed line for reference. Error estimates are not plotted here because they can not be represented properly on the logarithmic axis near the scale where the sign changes; the error estimates on these quantities can be seen in Figure \ref{fig:S3_fits}a,d.}
	\label{fig:S3}
\end{figure}

\begin{figure}
	\centering
	\includegraphics[width=1\linewidth]{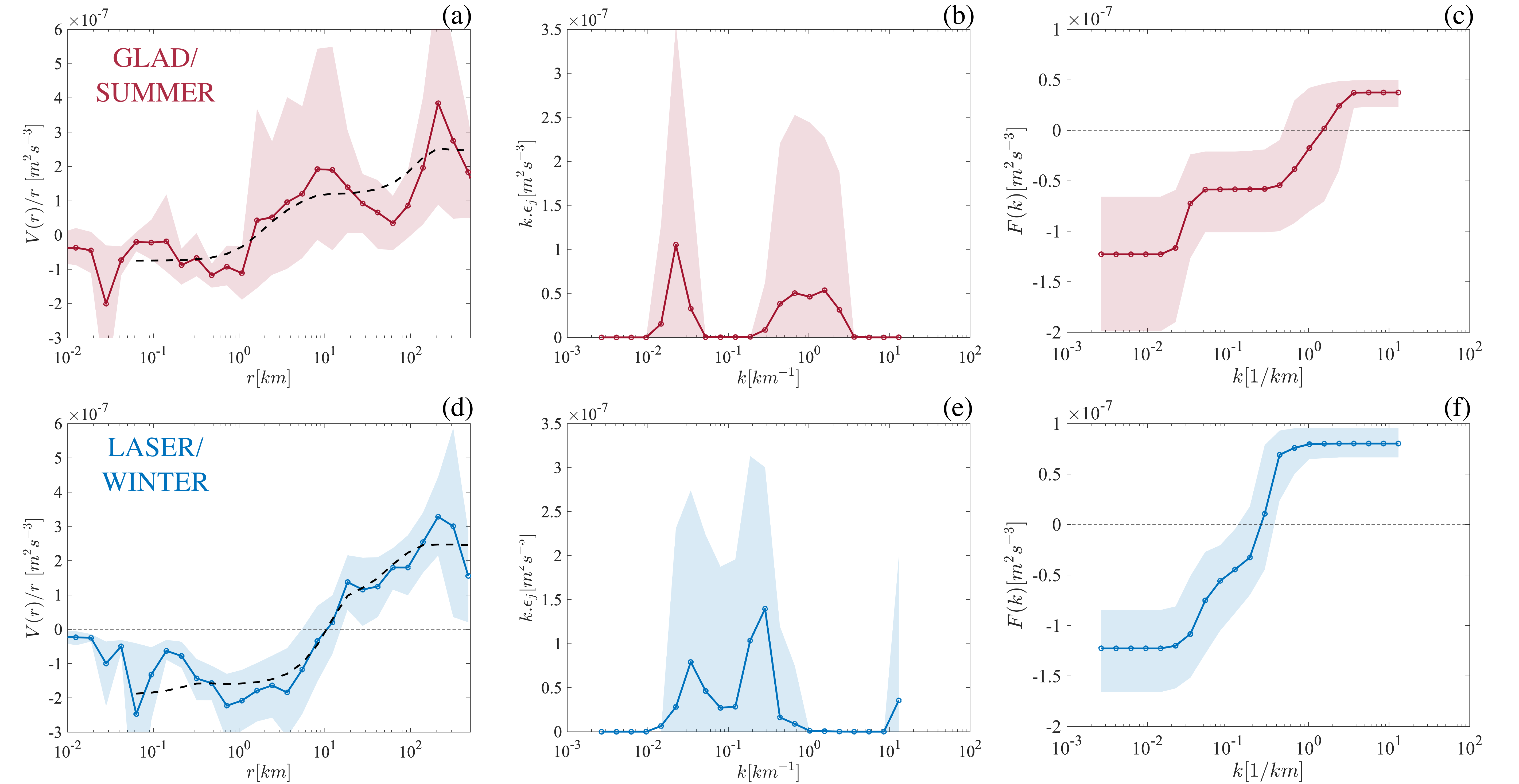}
	\caption{\textbf{Estimates of spectral flux and corresponding parameters from third-order structure functions for the GLAD/Summer (top; a-c) and LASER/Winter (bottom; d-f) experiments.} (a,d) The normalized third-order structure function ($V(r)/r$) and the fit, dashed black line, over scales of 50 m-500 km. (b,e) The estimated energy injection as a function of wavenumber, plotted in variance-preserving form to account for the logarithmic x-axis. Variance-preserving form accounts for the squeezing of the x-axis on a logarithmic scale and ensures that these panels can be visually interpreted such that the amount of energy injected corresponds to the area under the curve. (c,f) Spectral flux as a function of wavenumber. The positive and negative values denote upscale and downscale fluxes, respectively. %\com{fix axis ranges, maybe also show $\epsilon_u$ as a point or line in middle panels.}
	}
	\label{fig:S3_fits}
\end{figure}

The third-order velocity structure functions (SF3; $V(r) = \left< \delta u_L^3\right> + \left< \delta u_L \delta u_T^2 \right>$) are a metric that can be associated with the turbulent kinetic energy transfer rate \cite{Frisch1995}, and their sign, under certain hypothesis, is associated with the direction of the kinetic energy transfer -- with a negative SF3 indicating a forward or downscale transfer of energy and vice-versa. The GLAD experiment, conducted in the summer, provided the first observational evidence in the ocean that a wide range of scales, O($<$1km), had a negative SF3 \cite{Balwada2016, poje2017evidence, pearson2019impacts}, which is emblematic of a forward cascade of energy. 
The LASER experiment conducted a few years later in the winter season further solidified the generality of this observational result \cite{berta2020submesoscale, pearson2020biases}, and also showed a seasonal modulation in the length scale where the sign change happens (Figure \ref{fig:S3}). A similar result showing negative values of SF3 at smaller scales had been observed in the atmosphere about two decades earlier \cite{Cho2001}.

%The third-order structure function (SF3) from GLAD and LASER drifters have been presented in a number of previous studies \cite{Balwada2016, poje2017evidence, pearson2019impacts, berta2020submesoscale}, and have all interpreted the range of negative values at scales smaller than 1-10km as qualitatively emblematic of a forward energy cascade (Figure \ref{fig:S3}a,b). 

The interpretation, in these previous studies, of the direction of kinetic energy transfer based on the sign of the SF3 is rooted in classic inertial-range theories, e.g. Kolmogrov's 4/5th law \cite{Kolmogorov1941}. These interpretations are potentially suspect, or are at least only qualitatively correct, when inertial ranges cannot be clearly identified, or when the assumptions  used to reach inertial range arguments, e.g. purely two- or three-dimensional flow or asymptotic separation from forcing scales, are not satisfied. To overcome this major limitation we use a new theoretical framework developed by Xie and B\"uhler 2019 \cite{XieBuhler2019b} that allows us to directly infer the spectral fluxes from SF3, under the conventional assumptions of isotropy and homogeneity. In this framework, the spectral flux ($F(k)$) is defined by the corresponding energy injection rates ($\epsilon(k)$) at wavenumber $k$ (corresponding to scale $l=1/k$), and the upscale energy transfer rate ($\epsilon_u$), which are the parameters that represent the shape of $F(k)$. This spectral flux was analytically transformed to the corresponding SF3 in terms of the same parameters. The observational estimate of the SF3 can then be fitted using this analytical form, and all the parameters and thus the corresponding spectral flux can be inferred. 
To avoid amplifying small-scale error in the inversion process we made a physically motivated and pragmatic assumption that over the range of observed scales the energy injection is positive and only adds energy to the surface flow, which is equivalent to assuming that the spectral flux is an increasing function of wavenumber and that all the dissipation and extraction of kinetic energy out of the surface flow takes place outside the range of scales where the fitting is done. 
This is very well justified for dissipative mechanisms, which are active at scales much smaller than the ones observed here, but any transfer from kinetic energy to potential energy over the range of fitting scales has been ignored here (further details in Methods and SI). Even with this assumption the fitted SF3 matches the observed SF3 relatively well, capturing the broad structure within errorbars, and without fitting every small detail (Figure \ref{fig:S3_fits}a,c).

The estimated parameters provide us with two crucial physically relevant measures of the flow dynamics, the distribution of energy injection rate (Figure \ref{fig:S3_fits}b,c) and 
%the cumulative integral of the kinetic energy transfer function, which is equivalent to the energy injection rate under the assumption of statistically steadiness -- 
the spectral flux as a function of spatial scale or wavenumber (Figure \ref{fig:S3_fits}e,f). 
The energy injection shows two distinct peaks, one associated with smaller scales and another with larger scales. The large-scale peak is occurs around 40-50km, spreading between 20-100km, and does not vary significantly with the season. 
In contrast, the smaller-scale energy injection is modulated seasonally. In summer the energy injection peaks around 1km and is spread between 500m-5km, while in winter the peak is shifted to about 5km and is spread between 1km-10km. In summer the small- and large-scale peaks are distinct, while in winter the small scale energy injection is stronger in amplitude than summer and the small and large scale energy injections show a tendency to partially overlap. The scale where the spectral flux changes from being negative (upscale transfer) to positive (downscale transfer) is also approximately the same as the scale where the small-scale energy injection peaks, which is further correlated with the scale where the Rossby number starts to become $O(1)$ (Figure \ref{fig:S2}a inset). 

\begin{table}[h!]
\centering
%\begin{tabular}{ |c|c|c|c| } 
\begin{tabular}{ |p{0.14\textwidth}| p{0.26\textwidth}| p{0.27\textwidth}| p{0.27\textwidth}| }
 \hline
 Experiment/ Season & Upscale energy transfer rate, $\epsilon_u$ [$m^2/s^3$] & Large scale energy injection rate (10km -100km) [$m^2/s^3$] & Small scale energy injection rate (100m - 10km) [$m^2/s^3$]\\ 
 \hline
 LASER/ Winter & $1.23$ $(0.85-1.66) \times10^{-7}$ & $0.78$ $(0-1.36)\times10^{-7}$ & $1.25$ $(0.84 - 1.73)\times10^{-7} $  \\ 
 \hline
 GLAD/ SUMMER & $1.23$ $(0.66-1.99) \times10^{-7}$ & $0.64$ $(0-1.56) \times 10^{-7}$ & $0.96 $ $(0.53-1.41) \times 10^{-7}$ \\ 
 \hline
\end{tabular}
\caption{Estimated upscale energy transfer rate ($\epsilon_u$), large scale energy injection rate ($\int_{1/10km}^{1/100m} \epsilon_f(k) dk$), and small scale energy injection rate ($\int_{1/10km}^{1/100m} \epsilon_f(k) dk$). The values shown are the mean, and the range in the parenthesis are the 5$^{th}$ and 95$^{th}$ percentile.}
\end{table}

%Qualitatively the SF3 and the parameters estimated from it look similar between summer and winter, indicating that the non-linear transfers between scales are active year round (Figure \ref{fig:S3}). There are broadly two distinct ranges of scales at which energy is injected into the surface flow.
%The smaller scale lies roughly between 500m-10km and the larger scale is roughly between 20-100km. We interpret these scales, and to some degree distinguish these scales, based on dynamics associated with the mixed-layer deformation radius and the interior deformation radius. 
Quantitative estimates of the upscale energy transfer rate and energy injection rates are summarized in Table 1. The upscale energy transfer rate, the maximum negative value of the spectral flux (Figure \ref{fig:S3_fits}c,f), and the energy injection at the larger scales, is similar within errorbars between the two seasons. While the energy injected at the smaller scales is enhanced by about 25-30$\%$ in the winter. In summer about $75\%$ of the total injected energy, summed over both the large and small scale injections, is transferred upscale, while in winter this ratio is reduced to about $60\%$ - indicating a strengthening of the forward cascade. Note that in both summer and winter, a fraction of the energy injected at the smaller energy injection scale is transferred upscale.
Considering the almost unchanged upscale energy transfer rate (reflective of energy flux to scales greater than 100km), most of the extra energy injected in winter relative to the summer transfers downscale. %, without impacting the scales larger than the mixed-layer deformation radius. 

%There are quantitative seasonal changes that highlight how the seasonal differences in surface forcing and the response of the stratification to this forcing impacts the flow dynamics. In summer the total kinetic energy input into the surface flows is approximately $1.64\times 10^{-7} m^2/s^3$ and about 80$\%$ ($1.29\times 10^{-7} m^2/s^3$) of this energy cascades upscale. In winter the energy injection rate is enhanced by about 40$\%$, increasing to $2.29\times 10^{-7} m^2/s^3$, but only $65\%$ ($1.49\times 10^{-7} m^2/s^3$) of this is cascaded upscale. The rate of energy injection at the larger scale is similar between the two seasons, while the energy injection rate at the smaller scale likely corresponding to the mixed layer deformation is enhanced in the winter along with the range of scales where the peak injection takes places increasing in size, from 500m-2km in summer to 4km-8km in the winter. These energy injection rates and scales can also be represented as a spectral flux (Figure \ref{fig:S3}e,f), which similarly shows that in winter the forward cascade of energy is stronger, the scale at which the net cascade shifts from down to upscale is slightly larger, and the strength of the upscale cascade is also slightly enhanced. It is also worth noting that the scale at which the Rossby number becomes $O(1)$, indicating the scales where flows start to escape the control of rotation, also increases to slightly larger scales in winter (Figure \ref{fig:S2}c). 

The estimation of interscale energy transfers from the observed SF3 indicates that the range of scales from 100m-500km are energized primarily at two distinct scales, which are roughly analogous to the interior deformation radius, O(50km), and the mixed-layer deformation radius, O(1-10km). The energy injection near the mixed-layer deformation radius is seasonally modulated, with the injection rate and scale increasing slightly in winter. This is most likely due to a deepening of the mixed layer, which is in turn associated with more available potential energy for release and a larger scale of instability. About 25-40$\%$ of the energy injected into the system  cascades to smaller scales, forward cascade, at scales where the Ro$>$O(1), likely because the flow starts to escape the strong constrain imposed by rotation for energy to be transferred upscale. 
The flow in the winter has a higher propensity to cascade energy to smaller scales, which may be understood as the more energetic flow in winter having higher speeds and thus a greater likelihood to escape rotational effects.
These results provide the first direct estimate of the downscale energy transfer rate, $O(10^{-7}m^2/s^3)$, at the submesoscales. This estimate for the energy transfer rate is similar to those retreived for the energy dissipation in the mixed layer using microstructure estimates \cite{moum2001upper}, suggesting that the submesoscale flows can provide a direct pathway towards dissipation.

%The fitting finds that the dominant scale of energy injection is around $1km$, ranging from $500m$ to $2km$.
%These estimates parameters were used to estimate the corresponding spectral fluxes (Figure 2c), which shows an expected upscale cascade of energy at large scales and downscale cascade of energy at smaller scales. The spectral fluxes do not go to zero at the smallest and largest wavenumbers, as would be expected in a closed idealized steady turbulent system, because our domain is not closed and moreover because the theory does not account for the dissipation mechanisms that might be active in the range of observed scales. 
%On average, the total upscale energy transfer rate is $\epsilon_\mathrm{u}=8.4091\times 10^{-8} m^2s^{-3}$ and the total energy input rate is $\epsilon=\sum_{j=1}^{N}\epsilon_j=1.1394\times10^{-7}m^2s^{-3}$, so we can calculate $R=\epsilon_\mathrm{u}/\epsilon = 0.738$, indicating that around $3/4$ of the total injected energy transfers upscale. 
%A smaller energy injection is also seen between 10-100m, which could be a result of noise in the data or a signature of Langmuir turbulence. Estimating the properties of the energy flux is revealing but does not indicate the mechanisms and processes that result in it; we provide some speculative insight and interpretation of the mechanisms that can produce these fluxes in the next section.

%%%%%%%%%%%%%%%%%%%%%%%%%%%%%%%%%
%%%%%%%% Discussion %%%%%%%%%%%
%%%%%%%%%%%%%%%%%%%%%%%%%%%%%%%%%

\section*{Discussion}\label{sec_summary}

%% Results->conclusions
% Main discovery: evidence of dual cascade.
The observations of submesoscale turbulence at the surface ocean, in the Northern Gulf of Mexico, show the presence of a dual energy cascade; energy dominantly cascades towards small-scales at scales smaller than O (1-10 km) and towards large-scales at scales larger than O (1-10 km). The energy injection takes place primarily over two distinct scale ranges, at the smaller scales O(500m-10km) near the mixed-layer deformation radius and at the larger scales O(20-50km) near the interior deformation radius. The presence of a turbulent dual energy cascade has been hypothesised in the literature as a mechanism needed to accomplish the small-scale dissipation in the ocean. Intuitions on its existence stem from high-resolution ocean models and DNS of rotating-stratified flows, but this is the first time it has been directly confirmed in ocean observations using the estimates of third-order structure functions analyzed with an innovative methodology.  

% Seasonal cycle
The strength of the energy transfers as well as the scales at which the dual cascade develops are modulated seasonally, likely in response to the change in mixed-layer depth and strength of lateral buoyancy gradients set by the atmospheric forcing and freshwater river outflow. The net kinetic energy injection over the observed range of scales and the fraction of this energy cascading forward to smaller scales is enhanced in the winter, when the mixed layer is deeper. Even though the energy injected at small scales increases in the winter, this extra energy injection mainly leads to a strengthening of the downscale flux, without affecting the upscale energy flux at scales larger than 50km. 
The scale at which the net kinetic energy flux switches from being dominantly towards small-scale to dominantly towards large-scale seems to be correlated with the scale at which the local (in scale) Rossby number becomes O(1), the latter shifting towards larger scales in the winter as the net kinetic in the system increases. 

A seasonal modulation in the quasi-equilibrium distribution of kinetic energy over scales is also observed, as evidenced by second-order structure-function analysis and its Helmholtz decomposition. In winter there is more kinetic energy relative to the summer in the flow over the submesoscale range, and also this range of scales is dominantly composed of non-divergent motions. 
This suggests that the enhancement and changes in flow structures in the winter could likely be tied to a mechanism like mixed-layer instability, which strengthens in the presence of deeper mixed layers and would largely energize the balanced part of the flow. While in the summer, divergent motions  make a significant part of the kinetic energy over the submesoscale range, suggesting that the internal waves enhanced in amplitude by the shallower mixed layer or strain-driven ageostrophic frontogenesis are dominant.

% How our results generalize
Our analysis focused on observations from the Northern Gulf of Mexico, particularly the deeper ocean away from the continental shelf where the seasonal modulation in mixed layer depth is quite large. However, we believe that the presence of a dual energy cascade is likely to be a ubiquitous feature of submesoscale turbulence in the global ocean, and its particular properties, such as the scale at which the net flux changes sign, would be modulated by the local environment. Our results would indicate that if the $\Ro>O(1)$ at any scales in a region, then a forward cascade would likely ensue, which is possible as the dynamics would diverge from the traditional quasigeostrophic (QG) phenomenology. 
This is likely to be true everywhere in the surface ocean, supported based on results from high-resolution numerical simulations, e.g.,  \cite{torres2018partitioning}, which suggests that the submesoscale range of scales are more energetic than what QG dynamics would suggest in most places, while the particular dynamical contributions to these scales, balanced motions vs internal gravity waves, are modulated seasonally and spatially. 
It is worth noting that these high-resolution model simulations are far from being converged, and the departure from QG theory is likely to be even more stark as finer scales are resolved and in the real ocean.

% Note on key method improvements
This discovery of a well defined dual-energy cascade presented here rests on two key elements: high-density sampling of the surface flows using a dense array of surface drifters \cite{poje2014submesoscale, novelli2017biodegradable}, and theoretical advancement in structure-function analysis by generalizing the use of third-order structure functions \cite{XieBuhler2019b}. However, as most observational analyses go our results require assumptions and are likely to have some biases that are important to discuss, as they will chart the path for future research. The theory surrounding structure functions is developed in Eulerian coordinates and assuming homogeneity and isotropy, while the surface drifters sample the velocity field following the horizontal flow (sampling in a quasi-Lagrangian frame), with the real ocean having some degree of spatial inhomogeneity and anisotropy introduced by the complexity of the domain or forcing mechanisms. 
It has been shown previously for the GLAD and LASER data that the impacts of inhomogeneity and anisotropy are relatively weak, and do not affect the zeroth-order results \cite{wang2021anisotropic, Balwada2016}. The quasi-Lagrangian nature of the observations results in biased sampling, as the drifters have a tendency to cluster into convergent regions \cite{dasaro2018ocean, pearson2019impacts}. We think that this bias likely results in slightly overestimating the strength of the forward energy cascade at scales smaller than O(1km), as the convergent fronts are likely to be regions where the dissipative processes are stronger \cite{dasaro2011enhanced}. The exact quantitative impact of these biases remains an open area of research at the moment \cite{pearson2020biases}, and no methods to systematically correct for their influence are available.

% Going forward
Despite these caveats, direct observational confirmation of the dual energy cascade is a significant step forward in our understanding of how oceanic turbulence operates, and shows that a direct pathway for dissipation of mesoscale kinetic energy is present at the surface ocean. The observational evidence provided here is also relevant from a fundamental standpoint and supports a new paradigm in turbulence, that of a dual energy cascade developing in a fully three-dimensional fluid under the influence of rotation and stratification \cite{Pouquet2017, Marino2015}.
Our results and methods provide therefore important avenues for future research. 
On the dynamical side, it remains unclear what sets the ratio of the energy that is cascaded downscale vs upscale and how is this ratio seasonally modulated. One possibility is that it might be controlled by the strength of the internal waves, as shown in \cite{Marino2013,Marino2015,Xie2020}, which are more dominant in the summer. A decomposition of the third-order structure functions into contributions from different parts of the flow could be helpful to answer this question.
While a pathway for energy dissipation at the surface ocean has been confirmed by the present study, it remains to be understood what the relative strength of this pathway in dissipating energy and closing the ocean's energy budget is compared to mechanisms that are active in the interior, at land or ice boundaries, or via interaction with surface forcing.

One major advantage of our methodology is that it does not require the observational data sets to be gridded, and can easily handle observations with gaps and non-uniform sampling, which is the case for most observational platforms including satellites. Thus, it is also worth considering how our analysis methods can be expanded to other observational data sets in the future, such as the surface velocities obtained from the Global Drifter Program, measurements that will come from the Surface Water and Ocean Topography satellite, or more targeted observations from process studies where buoyancy measurements are made along with velocity and can allow for consideration of the kinetic and potential energy cycles simultaneously.

\section*{Methods}
\subsection*{The surface drifters}
All the data used in this study was collected by observations from surface drifters, which are instruments that follow the flow in the surface ocean and their locations are tracked using the Global Positioning System (GPS). 
The location information is then used to infer the surface velocities. The drifters used here tracked the flow in the top 60cm of the surface. Their positions were retrieved at 5min intervals with a nominal position error of $<10$m, and these raw position estimates were processed and provided by CARTHE as processed drifter trajectories, which were low pass filtered with a 1hour cutoff and resampled to a uniform time grid of 15 mins. Here we used drifters that were deployed in two targeted studies in the NorthEastern Gulf of Mexico.

The Grand LAgrangian Deployment (GLAD) experiment was conducted in the wake of the Deep Water Horizon oil spill. 297 CODE style surface drifters \cite{lumpkin2017advances} were released over the period of 11 days at the end of July 2012, making it the largest drifter release experiment at that time \cite{poje2014submesoscale}. The trajectories span the period from July to October 2012, which are the summer months. 

The LAgrangian Submesoscale ExpeRiment (LASER) was conducted at approximately the same location as GLAD. A total of approximately 1000 surface drifters were released near the end of January 2016, making it the largest drifter deployment to-date. The surface drifters used in LASER, CARTHE style drifters, had a slightly different design than the GLAD drifters, the new design was meant to be more rapidly deployable and also more environmentally friendly \cite{novelli2017biodegradable}. Despite the design differences the GLAD and LASER drifters showed similar characteristics at following the flow, and for all practical purposes are considered to be the same. The trajectories span the period from January to March 2016, which are the winter months. Many of the LASER drifters lost their drogue at some point after their deployment, and we only use the portion of the trajectories when the drogue was attached \cite{haza2018drogue}. 

The drifters were released in clusters, and so most of the observations at the smallest separation scales ($<10 days$) are for the duration of late July and early August in GLAD and late January and early February in LASER. The deployments were also often targeted on particular flow features, so the samples at the smallest scales might not be as representative of the true statistics as the samples gathered once the drifters disperse and randomly sample many different flow features. 

\subsection*{Statistical metrics and error estimates}
The metrics of interest in this study take the general form - 
$\left< \delta u^n \right>(r)$, where the $\left<.\right>$ indicates an ensemble averaging operation, $\delta u (r)$ is the difference in a particular velocity component between two points that are separated by distance $r$, and this velocity difference is raised to some power of $n$ corresponding to different orders of the structure functions.

When estimating these metrics using drifters we assume that the velocity measured by a particular drifter is the velocity of the surface ocean at that location. Using any two drifters we get one sample estimate of $\delta u (r)$. Since we want to estimate the metric as a function of separation scale we divided the separation axis, $r$, into bins that are logarithmically distributed between 10m and 1000km using the formula $r_n = r_0\times1.5^n$, where $r_0 =10m$ and $n=(0,1,2,3,...)$.
The ensemble averaging is replaced by collecting all pairs in any particular bin and time averaging over the entire duration of the experiment and space averaging, performed by averaging together all the scattered observations being collected by different drifters pairs; this is equivalent to assuming temporal stationarity and spatial homogeneity in a statistical sense. To ensure that the homogeneity assumption is not significantly violated by the samples, we only use pair samples collected by the drifters in a particular region that we expect to have similar dynamics everywhere (Figure \ref{fig:data_dist}). We also assume isotropy by averaging over all orientations of the position vectors connecting the two drifters relative to the geographical coordinates. 

The error estimates on the metrics are calculated by using a form of bootstrapping called modified block bootstrapping. Regular bootstrapping is done by estimating the same statistical metric multiple times by random sampling with replacement, keeping the number of samples the same as the original sample size, from the observed distribution of the samples (the samples for us are measurements of $\delta u(r)^n$ in some separation bin, where the number of samples is smaller than the number of drifter pairs). The mean over these multiple estimates of the metrics is then used as the estimate of the metric, and percentiles of these distributions can be used as estimates of the error - here we use the $5^{th}$ and $95^{th}$ percentiles. However, one key assumption in regular bootstrapping is that all the samples are independent, which is not even approximately true for our dataset because of the temporal and spatial correlations between the different pairs. Instead in block bootstrapping the dataset is divided into blocks that are approximately independent, and the resampling is done over these blocks rather than over all the individual samples. Here the blocks were defined by (a) estimating the total duration ($T_{tot}$) over which a significant number of data was collected, defined approximately as 90 days during GLAD and 60 days during LASER, (b) estimating the time scale corresponding to each scale ($T_{scale}(r)$), which was estimated as $T_{scale}(r) = r/\sqrt{D_{tot}}$, (c) defining the number of degrees of freedom as $N^{DOF} = T_{tot}/T_{scale}$, and then (d) dividing the total number of samples in each bin (arranged in order with the $m$ individual time series for the $m$ pairs that spent some time in the particular separation bin) into  $N^{DOF}$ blocks. This is an approximate but pragmatic procedure, and due to some degree of independence between the different pair time series in each bin actually results in an upper bound on the error bars, since we assume that there are less independent blocks than there actually might be (some independent time series from different pairs might end up in the same block when using our algorithm). This method is described further in section B of the Supp. Info.

\subsection*{$2^{nd}$ order structure functions as proxy for scale-wise kinetic energy distribution}\label{sec_2nd}
The longitudinal and transverse components of the second-order structure functions are defined as 
\begin{equation}\label{SF2_def}
    D_{LL} (r) = \left<(\delta u_L)^2\right>, \quad D_{TT} (r) = \left< (\delta u_T)^2 \right>
\end{equation}
where $\delta u_L$ and $\delta u_T$ are longitudinal and transverse velocity differences, respectively. These are defined as,
\begin{equation}
	\delta u_L = \delta\bu\cdot \frac{\bor}{|\bor|} \quad\mathrm{and}\quad \delta u_T = \delta\bu\cdot \boldsymbol{t},
\end{equation}
where $\boldsymbol{r}$ is the vector connecting the two points ($\boldsymbol{x}_1$ and $\boldsymbol{x}_2$) where simultaneous velocity observations are made ($\boldsymbol{r} = \boldsymbol{x}_2 - \boldsymbol{x}_1$), and $\boldsymbol{t}$ is the unit vector perpendicular to it ($\boldsymbol{r} \cdot \boldsymbol{t}=0$) on the horizontal plane. $\delta \bu = \bu_2 - \bu_1$ is the difference in velocity between the two points. We refer to the sum of these two components as the total second-order velocity structure function ($D_{tot} = D_{LL} + D_{TT}$). Plots of these components are presented in the supp info (Figure \ref{fig:SF2_components}).

The second-order structure function is related to the corresponding kinetic energy as an integral relationship, $D_i(r) = \int_0^{\infty} E_i(k) (1-J_0(kr)) dk$ with $i=LL$ or $TT$, which is a result of Fourier transformation of isotropic two-dimensional fields. Here $E_i(k)$ is the longitudinal or transverse component of the kinetic energy and $J_0(x)$ is the zeroth-order Bessel function (further discussion of this relationship can be found in \cite{Balwada2016}). This relationship implies that the SF2 is not precisely related to the kinetic energy at a particular scale, but is rather a weighted version of the kinetic energy distribution. Unfortunately, inverting the relationship to estimate the kinetic energy from the structure function results in an incoherent solution due to amplification of noise, and is therefore not attempted here. However, scalings observed in the second-order structure functions provide a very good measure of the distribution of kinetic energy as a function of scale, and the qualitative conclusions of this paper are robust. We confirm this by using an alternate metric of the distribution of kinetic energy as a function of scale, called the signature function (Supp info. section D), and also by testing the relationship between the structure function and kinetic energy using idealized functional forms with known kinetic distribution (not shown). 

The longitudinal and transverse components can be used to infer the rotational ($D_R$) and divergent ($D_D$) contributions to SF2, $D_{tot} = D_R + D_D$, using a Helmholtz decomposition \cite{Buhler2014, Lindborg2015}. This is performed by using the following formulae,
\begin{equation}
    D_R (r) = D_{TT} (r) + \int_0^\infty \frac{1}{r} (D_{TT} (r) - D_{LL} (r)) dr
\end{equation}
\begin{equation}
    D_D (r) = D_{LL} (r) - \int_0^\infty \frac{1}{r} (D_{TT} (r) - D_{LL} (r)) dr
\end{equation}
The rotational and divergent estimates are more useful as they are closely related to different dynamical regimes, where balanced vortical flows are primarily rotational in nature while ageostrohic flows or internal waves have a significant contribution from the divergent part. 
These formulae are derived under assumptions of isotropy and homogeneity. 
When these assumptions are not perfectly satisfied, the estimates can have some nonphysical characteristics:
In particular, at scales where one component (rotational or divergent) is significantly smaller than the other, the formulae can result in negative values for the second-order velocity structure functions (shown in Figure \ref{fig:SF2_decompose}, and can be seen as a range of scales where $SF2_D$ is not plotted in Figure \ref{fig:S2}). 
Some corrections can be introduced to account for factors like anisotropy, e.g. \cite{wang2021anisotropic}, but were not used here because for the particular datasets under consideration adding complexity would not impact the qualitative conclusions.  

\subsection*{$3^{rd}$ order structure functions and inter-scale energy transfers}\label{sec_3rd}
% Introduce V
The $3^{rd}$ order velocity structure function (SF3 or $V$), defined as,
\begin{equation}\label{V}
V = \left<{\delta u_L \br{\delta u_L^2 + \delta u_T^2}} \right>,
\end{equation}
is related to the spectral flux ($F(k)$) through a Fourier transform as,
\begin{equation}\label{V_F}
    V(r) = -4r \int_0^{\infty} \frac{1}{k} F(k) J_2 (kr)dk,
\end{equation} 
where $J_2(x)$ is the second-order Bessel function. 

Here we discretize $F(k)$ using a set of piecewise constant basis, 
\begin{equation}
    F(k) = - \epsilon_u +  \sum_{j=1}^{N_f} \epsilon_j  H (k - k_j) dk_j,
    \label{eqn:fk_discrete}
\end{equation}
where $\epsilon_u$ is the upscale energy transfer rate (units $L^2/T^3$), $\epsilon_j$ is the energy injection density (energy injection per unit wavenumber, units $L/T^3$) at wavenumber $k_j$. The total energy input into the system would be $\sum_j \epsilon_j  dk_j$, and this sum can be done over a fixed range of scales to estimate the energy input or injection (units $L^2/T^3$) at those scales (as done for Table 1). 

Substituting (\ref{eqn:fk_discrete}) into (\ref{V_F}), we obtain 
\begin{equation}
	V (r) = 2\epsilon_\mathrm{u} r - \sum_{j=1}^{N_f}4 \frac{\epsilon_j}{k_{j}}J_1(r k_{j}) dk_j, \label{eqn:V_n}
\end{equation}
which links the measurable third-order structure function with physically important quantities including the energy injection rates at different scales and the largest-scale upscale energy transfer rate.

Now, we estimate $V(r)$ from surface drifter data, and fit the structure function using (\ref{eqn:V_n}) to obtain $\epsilon_{u}$ and $\epsilon_j$, which are then be used to gain insights into the kinetic energy transfer across scales, e.g., by calculating the corresponding $F(k)$ using (\ref{eqn:fk_discrete}). Here we fit the estimated $V(r)$ over the range of scales from 50m-500km, to avoid scales where the estimates of $V(r)$ are less certain. However, the exact choice of this range did not impact the results of our study.
The details of this estimation problem (inverse problem) and the relationship between the spectral flux and SF3 are discussed in the supp info section F.

\bibliography{scibib}

\bibliographystyle{Science}

\section*{Acknowledgments}
\textbf{Funding:} DB gratefully acknowledges financial support from NSF grant OCE-1756882 and NASA award NNX16AJ35G. J-HX gratefully acknowledges financial support from the National Natural Science foundation of China grant No. 92052102. RM and FF gratefully acknowledge support from the project ``EVENTFUL" (ANR-20-CE30-0011), funded by the French ``Agence Nationale de la Recherche" - ANR through the program AAPG-2020.\\
\textbf{Author contributions:} DB was primarily responsible for the observational analysis, JHX for the theoretical input, and RM and FF for the numerical simulation input. All authors contributed to the interpretation of the results and writing of the manuscript. \\
\textbf{Competing interests:} The authors declare that they have no competing interests. \\
\textbf{Data and materials availability:} The surface drifter data is available at:\\ \url{https://data.gulfresearchinitiative.org/metadata/R1.x134.073:0004}. All the codes for the analysis and figure generation are available at \\ \url{https://github.com/dhruvbalwada/SF3_to_KEflux}.

%Here you should list the contents of your Supplementary Materials -- below is an example. 
%You should include a list of Supplementary figures, Tables, and any references that appear only in the SM. 
%Note that the reference numbering continues from the main text to the SM.
% In the example below, Refs. 4-10 were cited only in the SM.  

\newpage

%%%%%%%%%%%%%%%%%%%%%%%%%%%%%%%%%
%%%%%%%% Supp Info %%%%%%%%%%%
%%%%%%%%%%%%%%%%%%%%%%%%%%%%%%%%%
\part*{Supplementary Information}
\appendix
\section{GLAD and LASER data distribution}
Figure \ref{fig:data_dist} shows the distribution of data as a function of separation scale, time, and spatial location. A subset of the full data set (Figure \ref{fig:S2}) was chosen for the analysis, in region defined by 91W to 84W, North of 24N, and in ocean deeper than 500m, to ensure near homogeneity and similar dynamics over the samples. Lot of drifters during the LASER experiment lost their drogues, which leads to a rapid decrease of observed pairs in Figure \ref{fig:data_dist}e, even though approximately 3 times as many drifters as the GLAD experiment were deployed.

\renewcommand{\thefigure}{SI-\arabic{figure}}
\setcounter{figure}{0}
\begin{figure}[h]
	\centering
	\includegraphics[width=\textwidth]{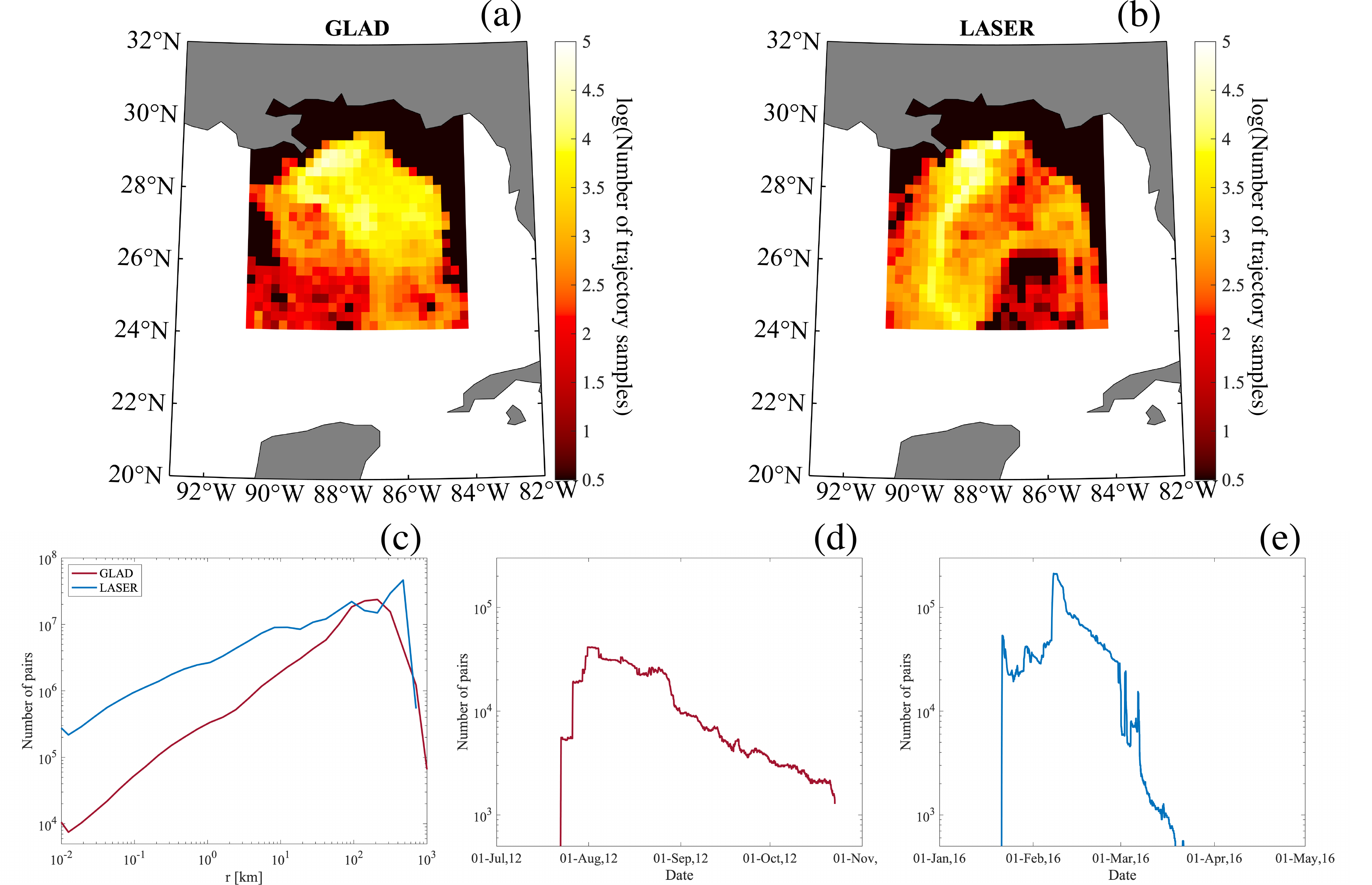}
	\caption{\textbf{Data distribution} (a,b) Number of trajectory samples in $0.25^o \times 0.25^o$ bins over the region that was used for the analysis (91W to 84W, North of 24N, and in ocean deeper than 500m), (c) number of pairs as per separation, (d, e) number of pairs as a function of time for GLAD and LASER respectively. }
	\label{fig:data_dist}
\end{figure}

\section{Error Estimates - Modified Block Bootstrapping}
\begin{figure}[h]
	\centering
	\includegraphics[width=\textwidth]{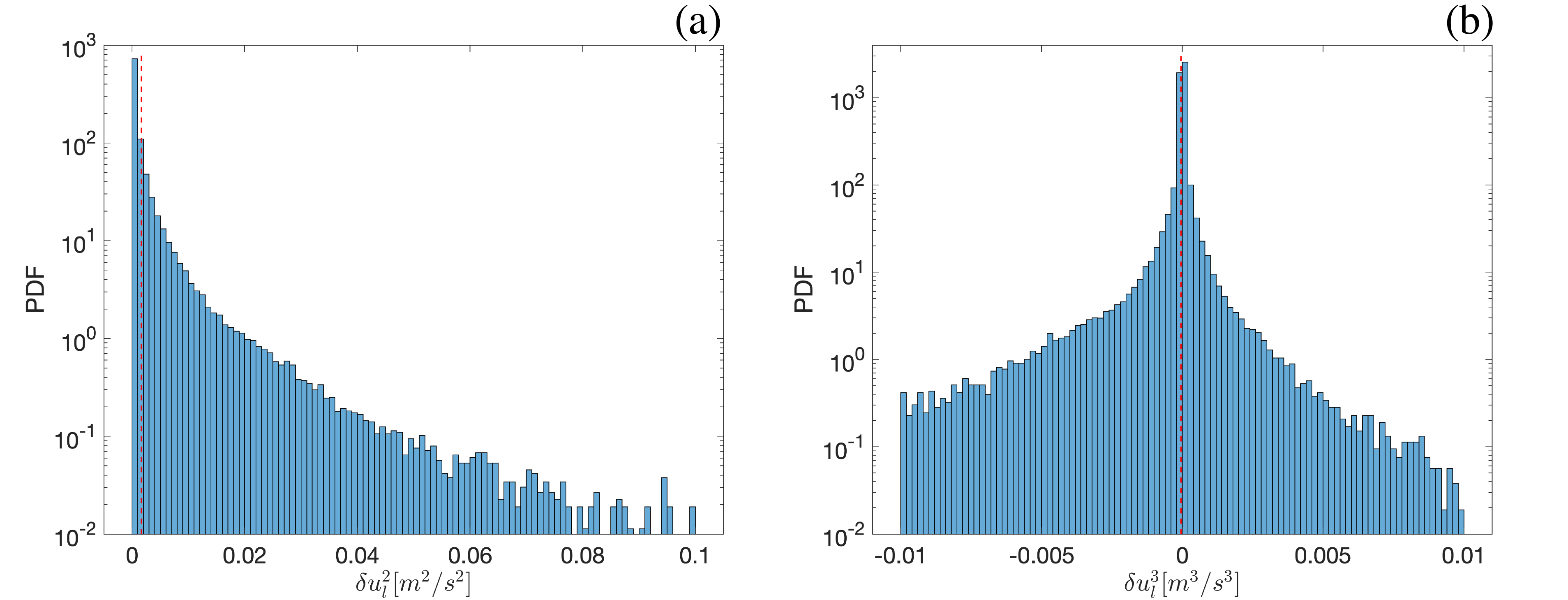}
	\caption{Distribution of $\delta u^2(r)$ (a) and $\delta u^3(r)$ for the GLAD experiment for the bin between 577 and 865m. The sample means are shown as the vertical red lines.}
	\label{fig:dun_dist}
\end{figure}

In this work we need to estimate errors on different statistical metrics that are all estimates of the mean, albeit of a complex looking statistic $\delta u(r) ^n$. Thus a method to estimate errors on the sample mean is needed. However, estimating this errors is non-trivial. Part of the reason for this complexity is that no analytical form is known for the sample distribution (distribution of $\delta u ^n$, Figure \ref{fig:dun_dist}), and so no standard formula for error is available. An alternative error estimation technique is bootstrapping (\url{https://en.wikipedia.org/wiki/Bootstrapping_(statistics)}), which involves random resampling with repetitions from the available data samples (mimicking the sampling process) and estimating the means using these randomly generated datasets. The distribution of these estimated means can then be used to infer the errorbars or the confidence intervals. 

\begin{figure}[h]
	\centering
	\includegraphics[width=\textwidth]{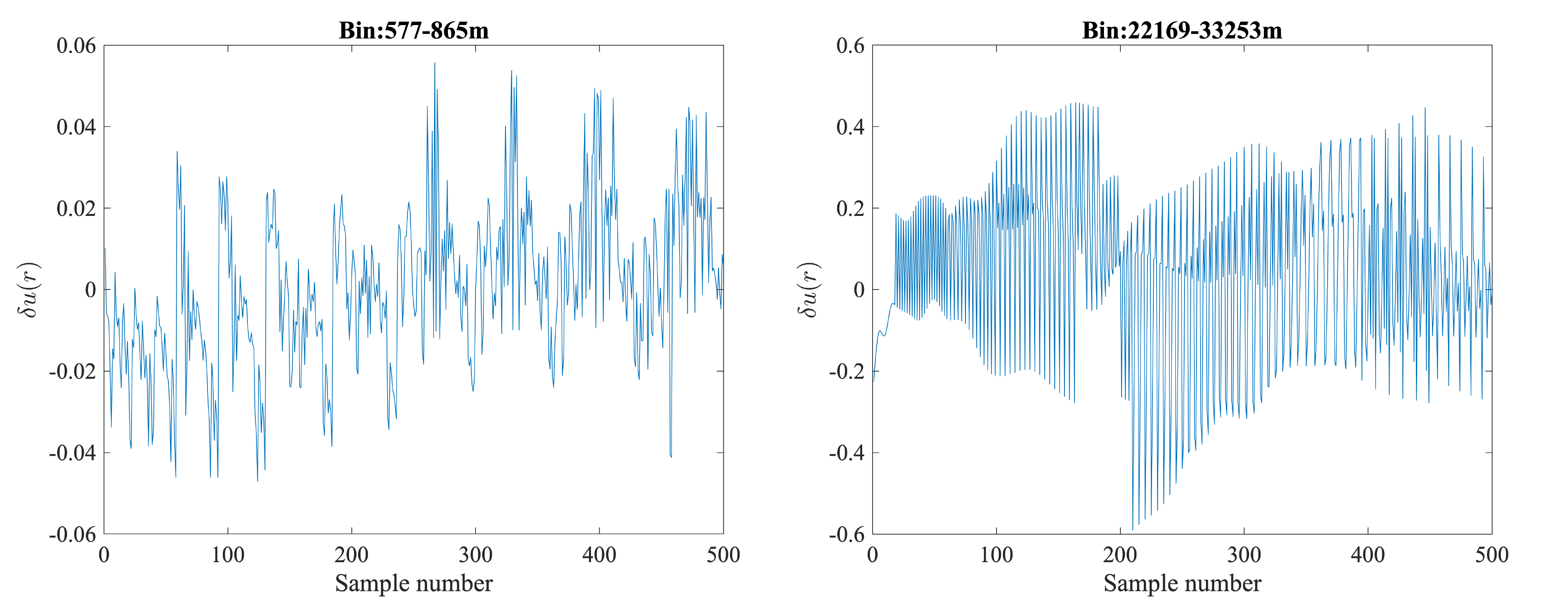}
	\caption{$delta u$ obtained from samples in two different separation bins -  showing temporal coherence, and hence a lack of independence between samples.}
	\label{fig:samples}
\end{figure}

This standard bootstrapping technique is not appropriate for us because it assumes that all available samples are independent, which is not the case here. Our samples are collected by pairs of surface drifters that are within a separation distance belonging to a finite-sized separation bin. Since the drifter pair can potentially stay in a particular bin for a finite amount of time, particularly in the larger bins at greater separations, we get a time series that will have some temporal correlation between samples (Figure \ref{fig:samples}). Additionally, if two or more pairs are in close proximity to each other and sampling the same flow feature the samples collected by the different pairs will also have some spatial correlation. The Lagrangian nature of the sampling mixes these temporal and spatial correlations. 

Since the samples are correlated we looked towards \textit{block bootstrapping}. This technique is usually used with time-series data, where the total time series is first divided into blocks based on some temporal correlation scale. For example, if the correlation scale is 10 days and the time series is 100 days long we would divide the timeseries into ten 10 day blocks. The resampling is then done using these blocks as individual samples. Here we do not have a single time series but rather a concatenated time series made up of different pairs that belong in a separation bin (Figure \ref{fig:samples}), where the different pairs are also likely to be correlated due to spatial proximity -- particularly at large-separation scales that are accounting for larger flow features that evolve slowly. Thus, using the traditional block bootstrapping also does not work for us; we experimented with this and found the error estimates to be intuitively too small (not shown). 

\begin{figure}[h]
	\centering
	\includegraphics[width=\textwidth]{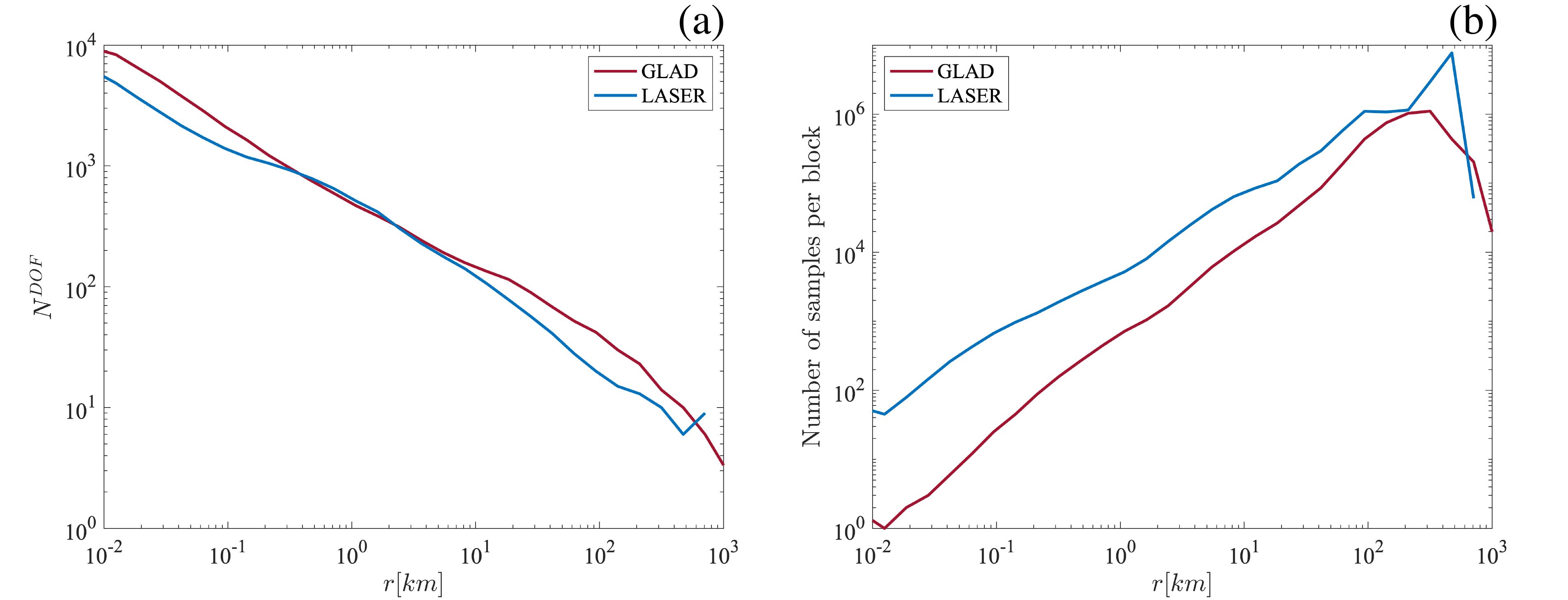}
	\caption{(a) Number of degrees of freedom (same as number of blocks) as a function of separation scale, and (b) number of samples per block as a function of separation scale.}
	\label{fig:blocks}
\end{figure}

Instead we use a slightly modified version of the block bootstrapping. We first estimate the possible degrees of freedom for any particular separation bin based on a temporal correlation scale and the total duration of the experiment. The temporal correlation scale was estimated using the estimate of the total SF2, as $T_{scale}(r) = r/\sqrt{D_{tot}(r)} $. The duration of the experiment, $T_{total}$ was chosen as roughly the duration over which a large number of pairs are present, and was chosen as $90$ days for GLAD and $60$ days for LASER (Figure \ref{fig:data_dist}d,e). Then the number of degrees of freedom was defined as $N^{DOF}(r) = T_{total}/T_{scale}(r)$, and is shown in Figure \ref{fig:blocks} for the two experiments. As one might expect the $N^{DOF}(r)$ decreases as a function of scale, since fewer independent large-scale events are sampled.
Then we took the concatenated set of samples for any separation bin and divided them into $N_{DOF}(r)$ blocks, and used these blocks as independent samples for block bootstrapping. 
Our underestimate of $N_{DOF}(r)$ is most likely smaller than the actually degrees of freedom (which is unknown), since we are not accounting for the fact that some samples might be uncorrelated because of large spatial separation, and is likely to give an upper bound on the error estimates. Our approach to estimating the blocks and consequently the errors is approximate but pragmatic, and we hope that future work can develop more precise error estimates. 

\section{Additional plots of Second-Order Structure Functions}
\begin{figure}[h]
	\centering
	\includegraphics[width=\textwidth]{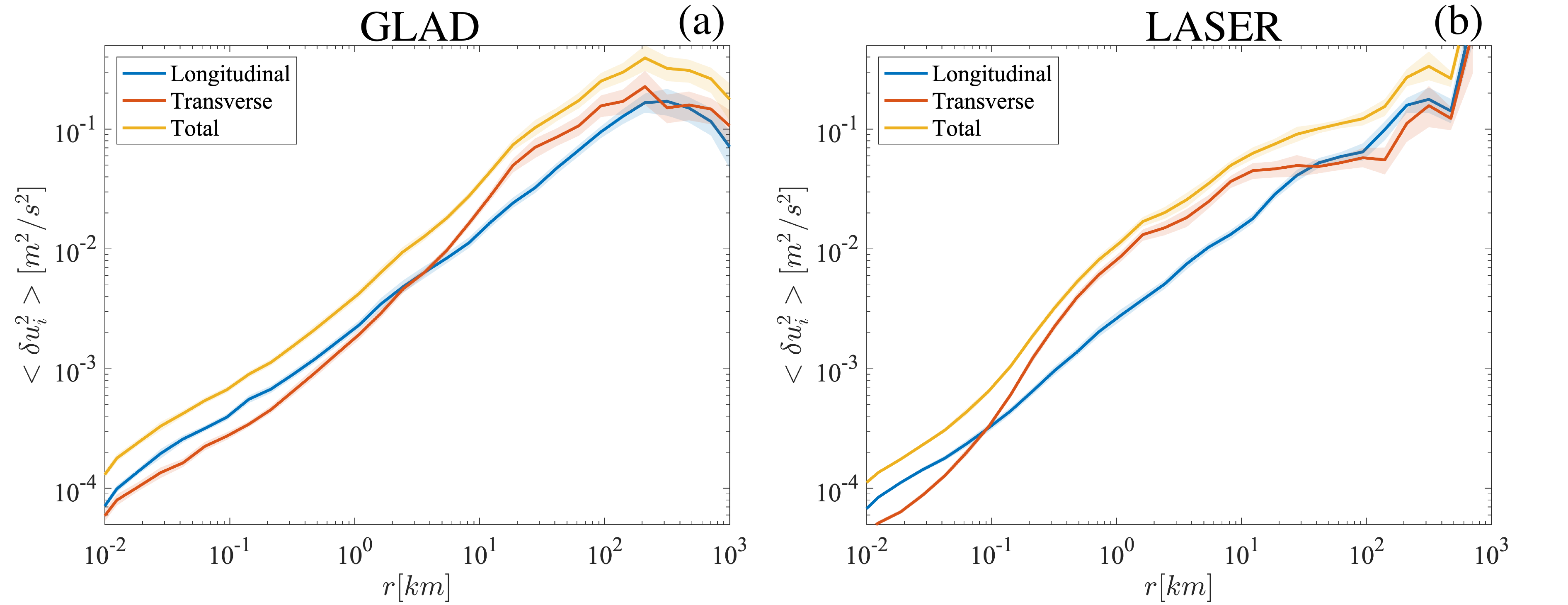}
	\caption{Longitudinal ($D_{LL}$) and transverse ($D_{TT}$) components of SF2 for the (a) GLAD  and (b) LASER experiments.}
	\label{fig:SF2_components}
\end{figure}

\begin{figure}[h]
	\centering
	\includegraphics[width=\textwidth]{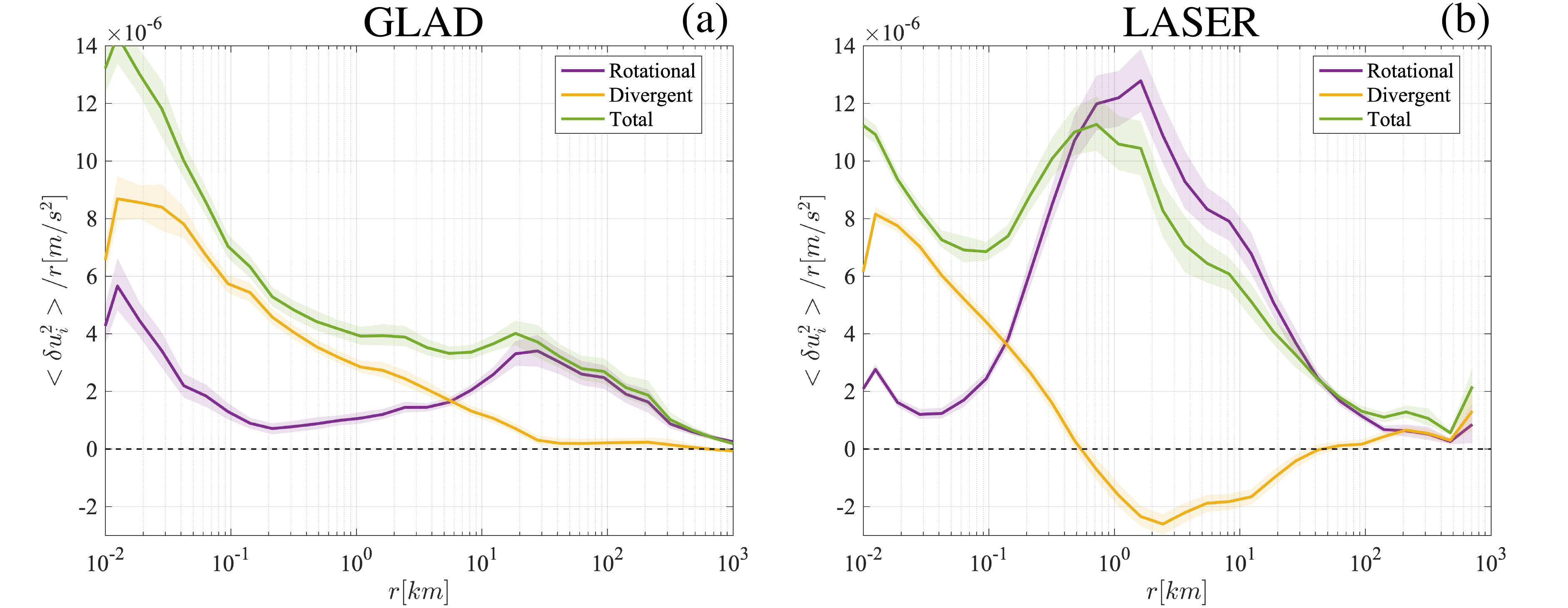}
	\caption{Rotational ($D_{R}$) and divergent ($D_{R}$) components of SF2 for the (a) GLAD  and (b) LASER experiments compensated by $r$ for visualization purpose on a linear axis.}
	\label{fig:SF2_decompose}
\end{figure}

Figure \ref{fig:SF2_components} shows the two raw components, longitudinal and transverse, of the total SF2 and figure \ref{fig:SF2_decompose} shows the two decomposed components, rotational and divergent, of the total SF2. The information in figure \ref{fig:SF2_decompose} is the same as figure \ref{fig:S2}, and is shown on a linear axis so that the negative values can be seen. In principle all the components of the SF2 should be positive, but the divergent components takes negative values over some range of scales. These negative values are small compared to the rotational or total components, about 20$\%$ at maximum but usually much smaller contribution to the total. This non-physical result is likely because the assumptions required to derive the structure function decomposition formulae (equation 3 and 4) are not perfectly satisfied, and this failure in the method has been discussed in length in \cite{wang2021anisotropic}. 

\section{Signature functions - an alternate metric for distributions of kinetic energy as a function of scale}

Second-order structure functions (SF2) represent a smoothed version of the distribution of kinetic energy across scales. 
This is because at any scale the SF2 includes a sum of energy from all smaller scales, but also enstrophy from all larger scales (which is often small as enstrophy is usually cascading downscale). 
\cite{davidson2015turbulence} proposed signature function as an alternative function that is better representative of the kinetic energy at different scales, and can be almost be considered comparable to the kinetic energy spectrum. It is defined as, 
\begin{equation}
G(r) = -\frac{r^2}{4}\frac{\partial}{\partial r}\br{\frac{1}{r}\frac{\partial}{\partial r} \left<\delta u^2\right>}.
\end{equation}

One caveat of the signature function when working with observational estimates is the requirement to estimate derivatives. Here we ameliorate the noisy nature of derivatives by fitting a 5th-order polynomial to the SF2 before estimating the signature function (the results here are not overly sensitive to this choice). 

\begin{figure}
	\centering
	\includegraphics[width=\textwidth]{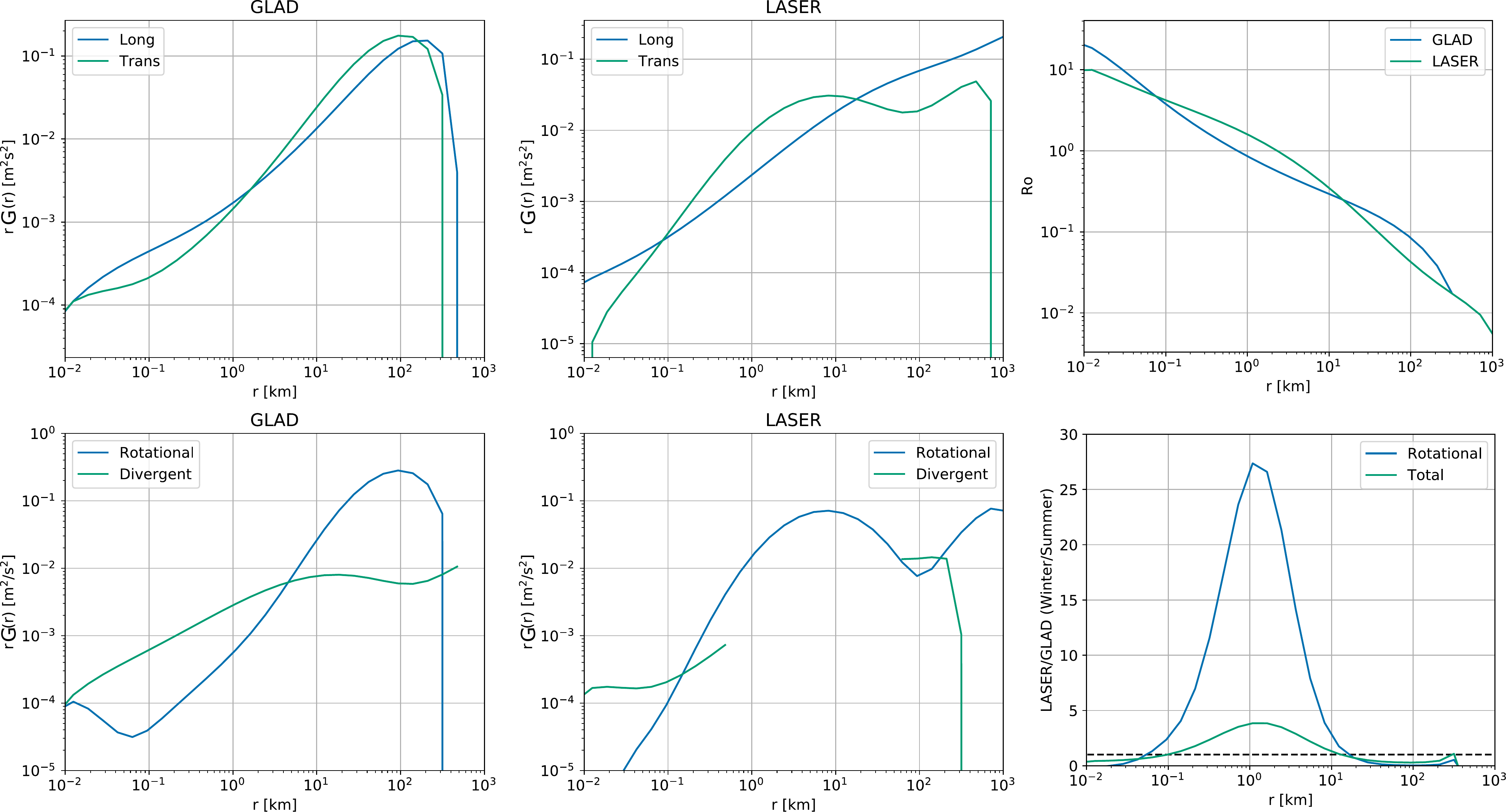}
	\caption{\textbf{Signature functions for the GLAD and LASER experiment.} (a, b) The longitudinal and transverse components. (c) the Rossby number for the two experiments. (d, e) The rotational and divergent decomposition. (f) The ratio of the total and rotational components between the two experiments.}
	\label{fig:signature_function}
\end{figure}

The signature function estimates are shown in Figure \ref{fig:signature_function}. While there are some quantitative differences compared to the SF2, in part due to the polynomial fitting, none of the major qualitative results about the relative behavior of winter vs summer or rotational vs divergent flows change. Part of the reason for this is that most of the enstrophy the type of flows we are considering is expected to be at the smaller scales, and so the SF2 is a good metric to express the distribution of KE across scales. %This is why we only present signature functions in the supp. info, also because the result is slightly sensitive to the order of the polynomial chosen for fitting. 

\section{Details of the third-order structure-function theory}
Here we first discuss the derivation of the Karman-Howarth-Monin (KHM) equation in our context, and then discuss the relationship between the third-order structure function and the spectral flux.

\subsection{Derivation of the KHM equation}
The goal of the theory is to describe the dynamics influencing the horizontal velocity correlations, and how the non-linear interactions leading to cross-scale energy transfers can be quantified by using the $3^{rd}$-order velocity structure functions. 
We focus on horizontal kinetic energy (KE), since in the ocean at the scales of interest the the horizontal kinetic energy is the dominant reservoir of KE, and also because the drifters only track the horizontal flow.

We consider the $f$-plane horizontal momentum equation,
\begin{equation}\label{mom_eq}
\bu_t + \nabla\cdot\br{\bu\bu} + \pa{z}\br{w\bu} + f \bu^\perp = -\nabla p + \bF +\bD,
\end{equation}
where $\bu=(u,\,v)$ is the horizontal velocity, $\bu^\perp=(-v,\,u)$, $w$ is the vertical velocity, $\nabla=(\pa{x},\,\pa{y})$ is the horizontal gradient, $f$ is the Coriolis frequency, $p$ is the pressure, $\bF$ is an external force, and $\bD$ denotes the dissipation. 
The dissipation term can contain both large- and small-scale dissipation.

To derive expressions for quantities that can be estimated using horizontal movement of drifters, we next consider two-point correlations with horizontal displacements.
We denote the horizontal locations of two points as $\bx_1$ and $\bx_2$, respectively, and $\bor=\bx_2-\bx_1$ is the horizontal displacement vector between these two points. 
Thus, we define the velocity at location $\bx_i$ as $\bu_i$, and the velocity difference is defined as
\begin{equation}
\delta \bu = \bu_2-\bu_1.
\end{equation}

By assuming \textit{horizontal homogeneity} we obtain the following relation between the spatial derivatives \cite{Frisch1995},
\begin{equation}
\nabla\equiv\nabla_{\bor} =\nabla_{\bx_2}=-\nabla_{\bx_1}.
\end{equation}
Homogeneity also results in the statistics of velocity difference to only be a function of the displacement. e.g. 
\begin{equation}
\ovl{\delta\bu^2}(\bx_1,\bx_2)=\ovl{\delta\bu^2}(\bor),
\end{equation}
where the $\ovl{(.)}$ denotes an ensemble average. Practically the ensemble average is approximated by a time and space average over all drifter pairs (discussed in ``Statistical metrics and error estimates" in the Methods). The assumption of homogeneity is foremost a pragmatic one, since it allows us to average over the full data set to ensure robust statistics. 

%Here, we assume homogeneity because we want to count the data of different drifters to make better statistics. Also, even for two specific drifters, since we cannot predetermine the location of drifters the homogeneity assumption makes the statistics reasonable. 

Multiplying $\bu_1$ to the momentum equation (\ref{mom_eq}) evaluated at $\bx_2$, adding the conjugate equation, and then assuming a statistically steady state, we obtain the steady Karman-Howarth-Monin (KHM) equation \cite{Frisch1995},
\begin{equation}
-\frac{1}{4}\nabla\cdot \bV = P + D, \label{KHM}
\end{equation}
where
\begin{subequations}
	\begin{align}
	\bV &= \ovl{\delta \bu |\delta \bu|^2}, \label{V}\\
	P &= \frac{1}{4}\nabla\cdot\br{\ovl{\bu_2|\bu_1|^2}-\ovl{\bu_1|\bu_2|^2}} - \frac{1}{2}\br{ \ovl{\bu_1\cdot\pa{z}\br{w_2\bu_2}} + \ovl{\bu_2\cdot\pa{z}\br{w_1\bu_1}} } \nonumber\\
		&\quad +\frac{1}{2}\nabla\cdot \br{\ovl{\bu_2 p_1}-\ovl{\bu_1 p_2}} + \frac{1}{2}\br{\ovl{\bu_1\cdot\bF_2}+\ovl{\bu_2\cdot\bF_1}} \label{P_exp}\\
	D &= \frac{1}{2}\br{\ovl{\bu_1\cdot\bD_2}+\ovl{\bu_2\cdot\bD_1}}.
	\end{align}
\end{subequations}
$\bV$ is the $3^{rd}$-order structure-function vector. $P$ subsumes terms corresponding to nonzero horizontal divergence, vertical velocity effect, pressure gradient and the external-forcing. 
%We treat $P$ as the energy-injection term because we can not estimate any of its contributions from observations; essentially equating energy injection to the contributions from external forcing and some flow dynamics. 
$D$ is the effect of dissipation.  
%do not have a direct measurement of $w$ and cannot calculate vertical derivatives based on the drifter data alone. 
%We want to use the $3^{rd}$ order structure-function vector $\bV$ to obtain the energy injection scale and the direction of energy transfer.

If we further assume hydrostatic balance in the vertical and vertical homogeneity we can express the pressure-related term in (\ref{P_exp}) through buoyancy $\theta$ as
\begin{equation}
    \nabla\cdot \br{\ovl{\bu_2 p_1}-\ovl{\bu_1 p_2}} = \overline{w_1\theta_2} + \overline{w_2\theta_1},
\end{equation}
which corresponds to conversion between potential and kinetic energy through processes such as baroclinic instability.

%\textbf{Is this term (first term on RHS of P) not present in the theory of 2D turbulence?
%	Does it go to zero under the isotropy assumption? 
%	Is there a way to write it as a $\delta$ u term? It might be measurable with drifters. The problem I can see is that it might be very sensitive to the choice of how the velocity is divided into the mean and the eddy part.
%	\com{Yes, this term does not present in the 2D turbulence where the horizontal velocity is incompressible, but here we cannot guarantee that. And I cannot see why this term will be zero when the flow is (horizontally) isotropic. e.g., similar term appear in the KHM equation of compressible turbulence. 
%	I cannot write it as a $\delta$ u term.
%	What's your definition of mean-eddy decomposition? 	
%}
%}
Further, we also assume \textit{isotropy}. \cite{Balwada2016, wang2021anisotropic} showed that isotropy is not strictly valid for the datasets under consideration, but it serves as a pragmatic assumption.
Also, we average measured structure functions in different directions with the same two-point distance ($r$), which is equivalent to an azimuthal/angle average that removes the dependence on orientation or angle \cite{Saffman1996,Nie1999,Casciola2003,Wan2009}.  
If the displacements of two measured points are uniformly distributed, this average procedure keeps the isotropic components and therefore the isotropic theory works.
Nuances associated with the assumptions of homogeneity and isotropy can be explored in future work. 

Thus,
\begin{equation}
\bV = V(r)\boldsymbol{e}_r,
\end{equation}
where $\boldsymbol{e}_r=(x/r,\,y/r)$ with $r=\sqrt{x^2+y^2}$ and $\boldsymbol{e}_r$ is an unit vector.
We can estimate $V$ directly from the velocity measured by the drifters,
\begin{equation}\label{V}
V = \ovl{\delta u_L \br{\delta u_L^2 + \delta u_T^2}},
\end{equation}
where $\delta u_L$ and $\delta u_T$ are longitudinal and transversal velocity differences, respectively. And they are defined as
\begin{equation}
	\delta u_L = \delta\bu\cdot \frac{\bor}{|\bor|} \quad\mathrm{and}\quad \delta u_T = \delta\bu\cdot \boldsymbol{t},
\end{equation}
where the unit vector $\boldsymbol{t}$ satisfies $\boldsymbol{t}\cdot \bor = 0$ and $\boldsymbol{r}\times\boldsymbol{t}=\boldsymbol{z}$ with $\boldsymbol{z}$ the vertical unit vector.
 
\subsection{Relationship between the third-order structure function and spectral flux}
We showed above that the the SF3 or $V(r)$ is present as a term in the KHM equation, which describes the dynamics of two-point velocity correlations. An analogous equation to the KHM equation can also be derived in spectral space, which describes the dynamics of the velocity power spectra \cite{Frisch1995}. 

Note that the two-point correlation and power spectra are related via a Fourier transform.
Thus, the third-order structure function can be related to the spectral flux via an integral relationship, 
\begin{equation}
    V (r) = -4r \int_0^{\infty} \frac{1}{K}F(K) J_2 (Kr) dK,
    \label{eqn:V2F}
\end{equation}
where $J_2$ is the second-order Bessel function. The Bessel function emerges from a Fourier transform due to the assumption of isotropy \cite{XieBuhler2019b}. 

\subsubsection*{Case of a single injection scale}
A common ideal scenario that is considered in turbulence theories is an energy injection at a single forcing scale, which corresponds to the spectral flux
\begin{equation}
    F(K) = -\epsilon_u + \epsilon H(K-k_f),
    \label{eqn:F1}
\end{equation}
where $\epsilon_u$ is the energy flux upscale and $\epsilon$ is the energy injection at the scale corresponding to $k_f$. This expression assumes that some fraction of the injected energy is fluxed upscale and the rest downscale, here the magnitude of downscale energy flux is $\epsilon_\mathrm{d}=\epsilon-\epsilon_\mathrm{u}$. This is regarded as an ideal case as the dissipation only happens at zero and infinite wavenumbers.

Substituting (\ref{eqn:F1}) into (\ref{eqn:V2F}) results in the corresponding expression of the third-order structure function \cite{XieBuhler2019b}:
\begin{equation}\label{V_bi}
\begin{aligned}
V = 2\epsilon_\mathrm{u} r - 4 {\epsilon}{k_f}J_1(k_f r),
\end{aligned}
\end{equation}
where $J_1$ is the Bessel function of the first order.

Expression (\ref{V_bi}) has three power-law ranges, corresponding to the inertial ranges of upscale energy transfer, downscale energy transfer and downscale enstrophy transfer, respectively. These ranges are shown in the asymptotic expansions as follows
\begin{equation}\label{V_expan}
V =\left\{\begin{matrix}
\underset{\mathrm{downscale\,energy}}{\underbrace{-2\epsilon_\mathrm{d} r}} + \underset{\mathrm{``enstrophy"}}{\underbrace{\dfrac{1}{4} \br{\dfrac{r}{l_f}}^2\epsilon r}} + O\br{\br{\dfrac{r}{l_f}}^5}, \quad \mathrm{when} \quad \dfrac{r}{l_f} \ll 1, \\
\underset{\mathrm{upscale\,energy}}{\underbrace{2\epsilon_u r}} + O\br{\br{\dfrac{r}{l_f}}^{-1/2}}, \quad \mathrm{when} \quad \dfrac{r}{l_f}\gg 1.
\end{matrix}\right.
\end{equation}
% The general application for V, which we will be using. 
By capturing the three inertial ranges in one formula with resolved forcing scale and with the applicability to the scenario of bidirectional energy transfer, expression (\ref{V_bi}) solves shortcomings of the previous theories which are only applicable to inertial ranges with unidirectional energy transfer, and therefore it provides the foundation of analyzing third-order structure function measured in geophysical turbulence.
Another advantage of the new theory is that it can be used to detect the energy injection scales and the magnitude of energy injection at each scale, which was not possible using classic inertial-range theories because inertial range by definition is away from the forcing scales.

\section{Details of parameter estimation method}
\subsection{Formulation of the discrete problem}
Gaining inspiration from equation \ref{eqn:F1}, we can express any general form of the spectral flux as 
\begin{equation}
    F(k) = - \epsilon_u +  \sum_{j=1}^{N_f} \epsilon_j H (k - k_j) dk_j,
\end{equation}
where $\epsilon_u$ is the upscale energy transfer rate (units $L^2/T^3$), $\epsilon_j$ is the energy injection density at scale $k_j$ (energy injection per unit wavenumber, units $L/T^3$).
Note that we indexed $dk_j$ to denote that the forcing wavenumber spacing does not need to be regular. This equation is a discrete representation of the true spectral flux using a set of piece-wise constant basis function.

This equation above can then be passed through the same procedure as what was done for the single forcing scale (equation \ref{V_bi}), to derive the corresponding expression for the $3^{rd}$-order structure function ($V$),
\begin{equation}
	V (r) = 2\epsilon_\mathrm{u} r - \sum_{j=1}^{N_f}4 \frac{\epsilon_j}{k_{j}}J_1(r k_{j})dk_j. \label{eqn:V_n}
\end{equation}
We will fit the observational estimate of SF3 using this expression to estimate the parameters, and hence obtain an estimate of the corresponding spectral flux. 
We do not directly use (\ref{V_F}) to obtain energy flux from the observed third-order structure functions to avoid the amplification of small-scale error in the inverse problem.

The V is estimated at discrete scales $r_i$ with $i=1,\,2,\,...,\,N_r$, which are set based on the used binning. 
%We can pick scales $l_{fj}$ with $j=1,\,2,\,...,\,N_f$ for energy injection. 
Thus, we obtain a linear equation for $\epsilon_u$ and $\epsilon_j ( = \epsilon_f(k_j))$,
\begin{equation}\label{eqn:big_mat}
    \begin{bmatrix}
    V(r_1) \\
    V(r_2) \\
    ...\\
    V(r_{N_r}) 
    \end{bmatrix}
    =
    \begin{bmatrix}
    2 r_1 & - 4\frac{dk_{1}}{k_1}J_1(r_1 k_{1}) & - 4\frac{dk_{2}}{k_2}J_1(r_1 k_{2}) & ... & - 4\frac{dk_{N_f}}{k_{N_f}}J_1(r_1 k_{N_f}) \\
    2 r_2 & - 4\frac{dk_{1}}{k_1}J_1(r_2 k_{1}) & - 4\frac{dk_{2}}{k_2}J_1(r_2 k_{2}) & ... & - 4\frac{dk_{N_f}}{k_{N_f}}J_1(r_2 k_{N_f}) \\
    ......\\
    2 r_{N_r} & - 4\frac{dk_{1}}{k_1}J_1(r_{N_r} k_{1}) & - 4\frac{dk_{2}}{k_2}J_1(r_{N_r} k_{2}) & ... & - 4\frac{dk_{N_f}}{k_{N_f}}J_1(r_{N_r} k_{N_f}) \\
    \end{bmatrix}
    \begin{bmatrix}
    \epsilon_u \\
    \epsilon_1 \\
    \epsilon_2 \\
    ...\\
    \epsilon_{Nf}
    \end{bmatrix}.
\end{equation}
The large matrix on the RHS with the unknown parameters has size $N_r\times (N_f+1)$. Thus the problem becomes one of solving a linear system of equations, 
\begin{equation}
    \mathbf{A} \mathbf{x} = \mathbf{b},
    \label{eqn:Linsys}
\end{equation}
where $\mathbf{x}$ is the vector of unknown $\epsilon$s, $\mathbf{b}$ is the vector of observed SF3, and the $\mathbf{A}$ is the matrix formed by the particular relationship (equation \ref{eqn:big_mat}) between them. 
Solving such system of equations is the subject of discrete inverse theory. Since we get to set $N_r$ and $N_f$ we can setup this problem as an under-determined ($N_r<N_f+1$), even determined ($N_r=N_f+1$), or over-determined (($N_r>N_f+1$) problem. Here we decided to take the over-determined route, and find the solution using a form of the least-squares method. 

\subsection{Least-squares method}
\begin{figure}
	\centering
	\includegraphics[width=\textwidth]{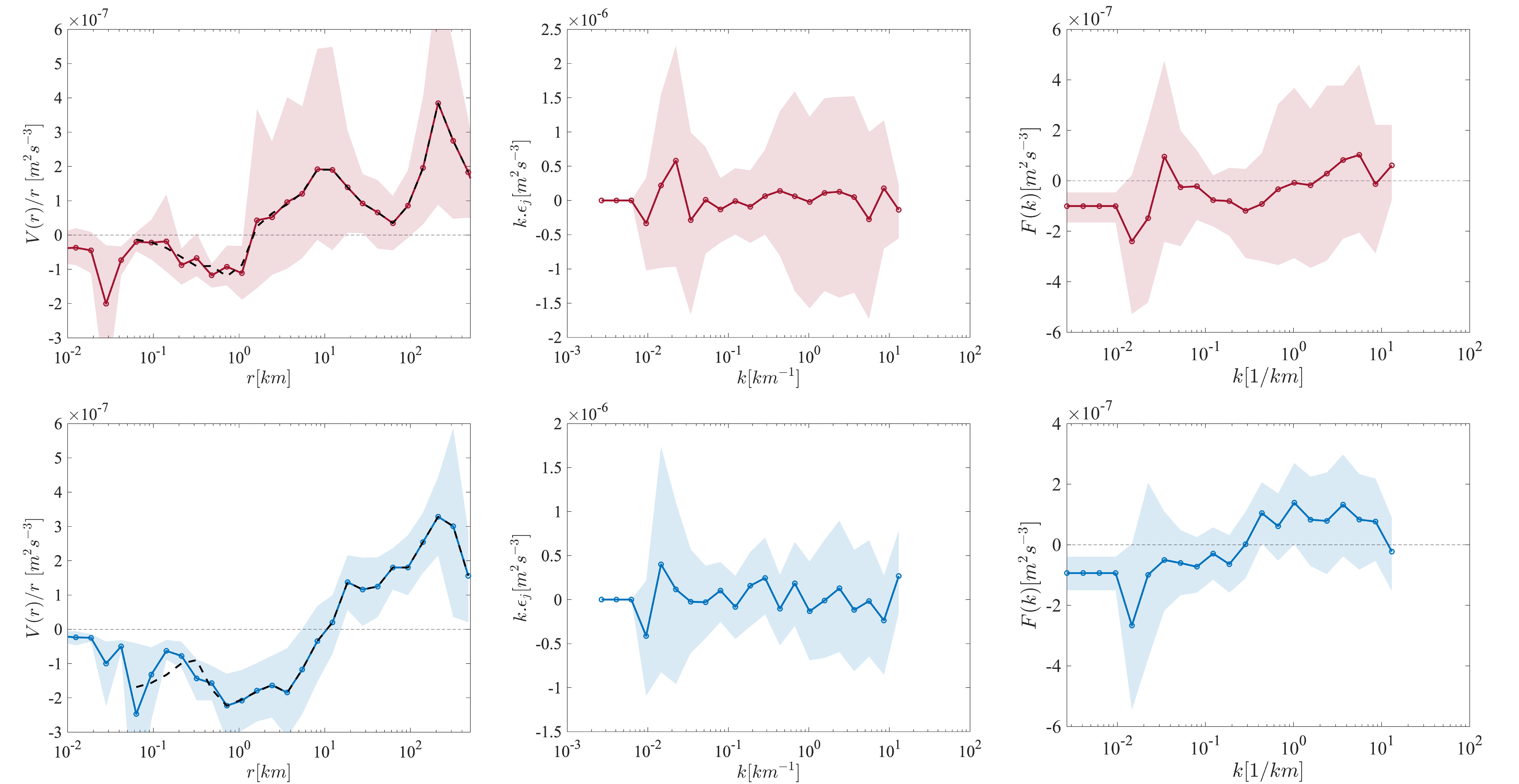}
	\caption{Fits to V(r) (dashed black line in first column) and parameter estimates using the least-square method for the GLAD (top) and the LASER (bottom) experiments. }
	\label{fig:fits_LS}
\end{figure}

The least-squares method solves equation \ref{eqn:Linsys} by minimizing,
\begin{equation}
    ||\mathbf{Ax} - \mathbf{b}||^2_2,
    \label{eqn:min}
\end{equation}
the L-2 norm (mean square error). 
If we solve this optimization problem in MATLAB drectly using the $\backslash$ operator, we obtain solutions presented in Figure \ref{fig:fits_LS}. 

The fitted $V(r)$ matches the observed $V(r)$ really well, capturing most of the details. The inferred energy injection is very variable and it is impossible to derive much physical insight from this. The estimated spectral flux is also quite variable, but a rough pattern of downscale flux at smaller scales and upscale flux at larger wavenumbers is suggested. 

The problem with this solution is that the optimization method has over-fit the data, providing a really good fit to every detail but little insight. This problem is addressed by using regularization, where some penalty term is added to equation \ref{eqn:min} to impose some additional physical constraints (like smoothness). The regularized solution can be designed to not all the stochastic variability in the observations, but instead only the broader pattern that might be more physically interpretable. One particular regularization approach that we used is discussed next.

\subsection{Non-negative least-squares method (used here)}
Here we use a particular type of regularization where it is assumed that all the parameters to be estimated are non-negative ($\mathbf{x} \geq 0$). Thus, we also assumed that $\epsilon_u$ and $\epsilon_j$ are all positive, and used the function \textit{lsqnonneg} in MATLAB to solve the system (\ref{eqn:Linsys}). This assumption is equivalent to assuming that the spectral flux, $F$, is an increasing function. 
This is physically justified because we expect there to be an inverse cascade at smaller wavenumbers followed by a forward cascade at larger wavenumbers, and also expect the dissipation to take place at scales outside the observed range. The one downside of this assumption is there is some sink of kinetic energy over the fitted scales, like conversion of surface kinetic energy to potential energy, it will be artificially smoothed over. However as shown in the previous section, some regularization is necessary to derive more physical insight and so we chose this pragmatic approach. The solution from this method is shown in Figure \ref{fig:S3_fits} and discussed in the main text.
In future work other regularization methods, like constraints only on smoothness, can be tried in the fitting procedure.

\section{Application to a simulation of rotating stratified turbulence}\label{sec_num}

To show the efficacy of the above methodology at detecting the energy injection scales and the spectral flux, we first apply it to a numerical simulation of rotating stratified turbulence. The numerical simulation is of a 3D triply-periodic incompressible boussinesq equations, and were presented in \cite{Marino2015}. A mean constant stratification is prescribed. The external mechanical forcing is isotropic and generated randomly, applied in a shell of modes with wave numbers ($k_f=10$). 

\begin{figure}
	\centering
	\includegraphics[width=0.32\linewidth]{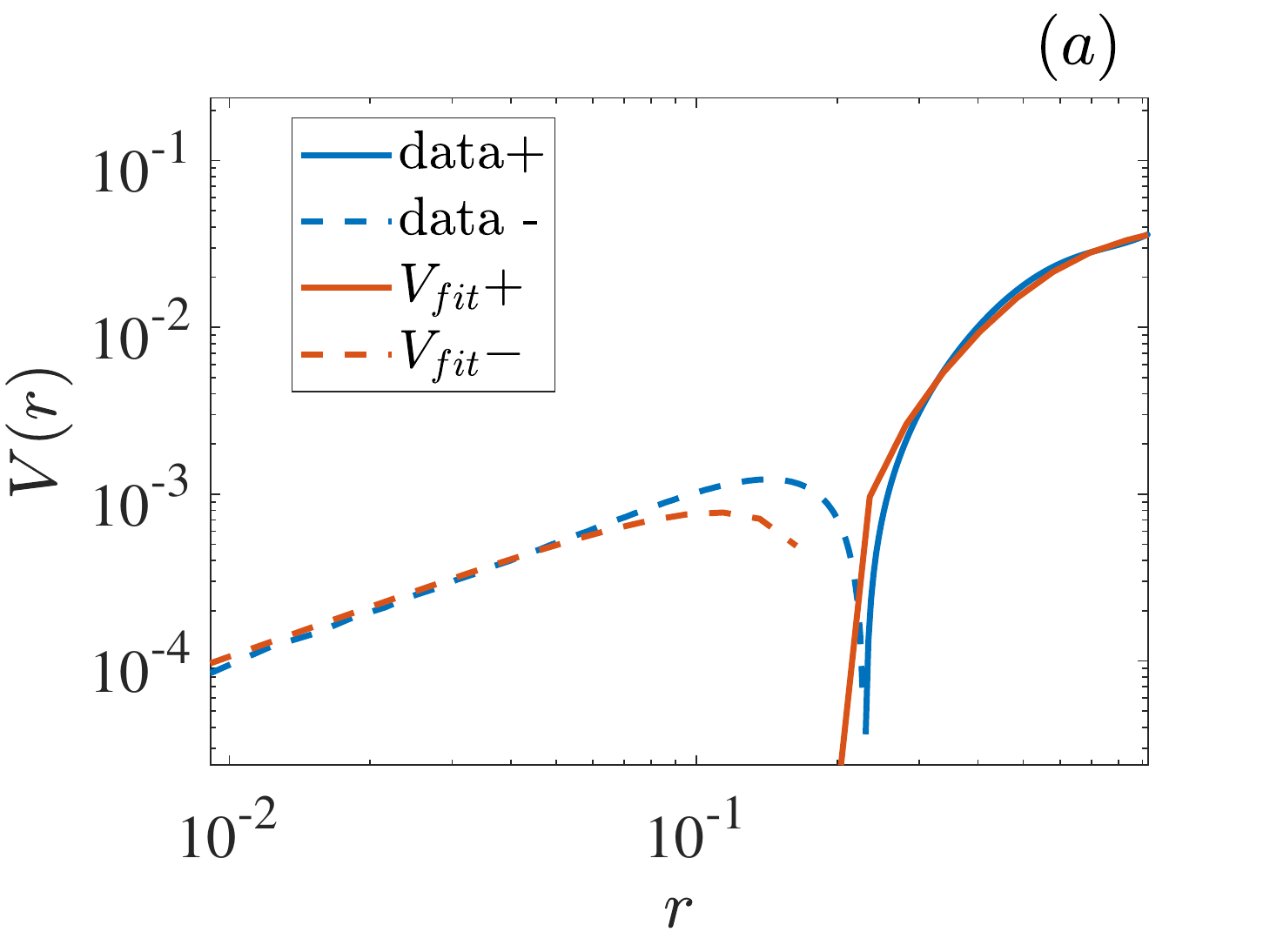}
	\includegraphics[width=0.32\linewidth]{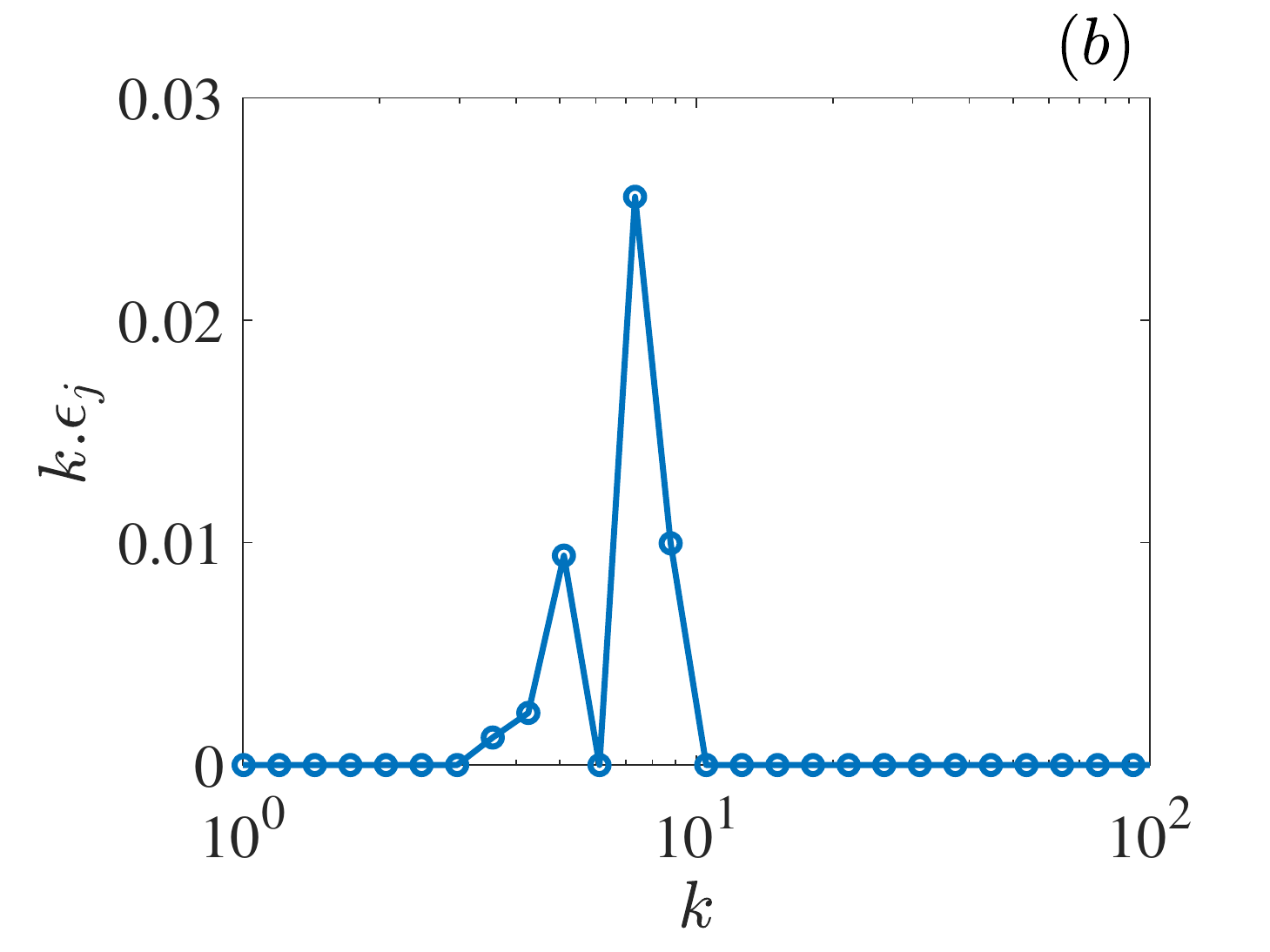}
	\includegraphics[width=0.32\linewidth]{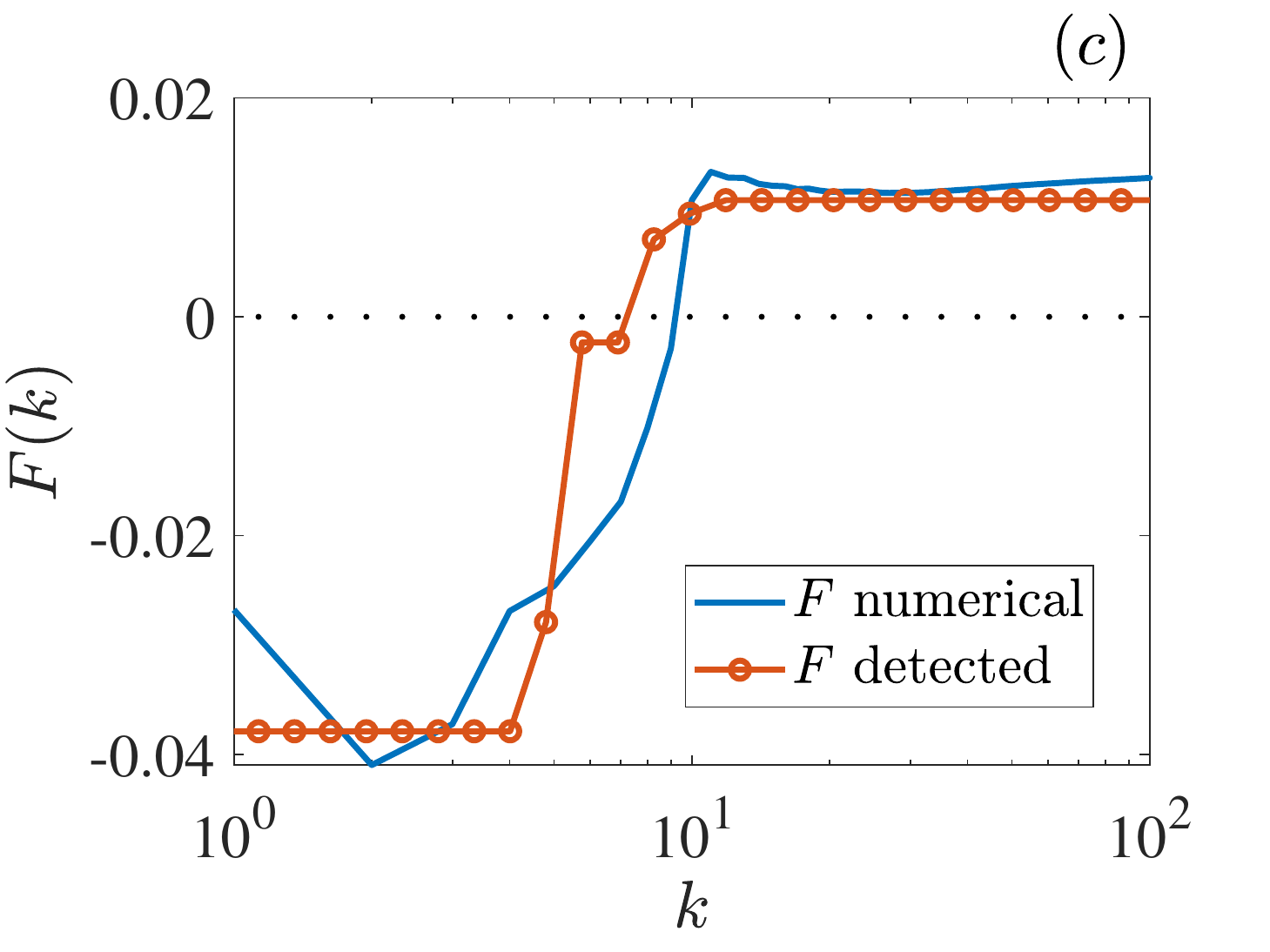}
	\caption{Fitting of the kinetic third-order structure function, the detected energy input and energy flux.}
	\label{fig_V_EK}
\end{figure}

This simulation was selected because it has the dynamical components that we expect to see in the ocean, we have precise knowledge of the energy injection, and a large range of scales smaller than the forcing scale are resolved and simulate a downscale energy transfers. 
%In fact, a snapshot of vorticity (Figure \ref{fig_V_EK}a) looks similar to the elements we see in the highest resolution ocean models. 
Also, more practically the qualitative structure of the $V$ is similar to what is observed and shown in Figure \ref{fig:S3}; $V$ is negative at smaller scales and positive at larger scales, and its absolute value approximately follows a linear power law (Figure \ref{fig_V_EK}a). We show that equation \ref{eqn:V_n} can be fit quite well (red line in Figure \ref{fig_V_EK}a) to the model $V$, by optimizing the free parameters: the energy injection rates and scales (Figure \ref{fig_V_EK}b) and the upscale energy flux. 
 
Furthermore, using the detected energy injection rate and the upscale energy flux we can reconstruct the kinetic energy flux in the spectral space, which is compared with that obtained directly from the numerical simulation in Figure \ref{fig_V_EK}c. Notice that the fitting only approximately matches the numerical $V(r)$ over the range of scales where it is negative, this is because $F(k)$ is not a perfectly monotonically increasing function. Implying that some small negative values of $\epsilon_j$ would be needed for a better fit, as should be expected given the slight decrease in F(k) at wavenumbers larger than 10. In fact, these scales are associated with a transfer from kinetic to potential energy. Our current method can not detect this detail. However, given the large uncertainty range associated with the observational $V(r)$, we should not expect to capture this level of detail even if we used an alternate method. The satisfactory fitting using our method implies that obtaining energy flux information from the third-order structure function is possible. 

%Also, in this controlled numerical simulation, the forcing wavenumber is $k_f=10$, which is well captured by our method. 
%\begin{figure}
%	\centering
%	\includegraphics[width=\textwidth]{./figures/figure_s1.pdf}
%	\caption{(a) A horizontal slice of vertical vorticity from the simulation. (b) Fitting of the kinetic energy third-order structure function $V$, (c) Comparison between the numerically obtained and detected horizontal kinetic energy flux, and (d) the detected energy injection rate at different scales.}
%	\label{fig_V_EK}
%\end{figure}

\end{document}